\documentclass[10pt,letterpaper]{article}
\usepackage[top=0.85in,left=2.75in,footskip=0.75in]{geometry}
\usepackage{amsmath,amssymb}
\usepackage{ragged2e}
\usepackage{changepage}
\usepackage{textcomp,marvosym}
\usepackage{cite}
\usepackage{nameref,hyperref}
\usepackage[right]{lineno}
\usepackage[nopatch=eqnum]{microtype}
\DisableLigatures[f]{encoding = *, family = * }
\usepackage[table]{xcolor}
\usepackage{array}
\usepackage{threeparttable}
\usepackage{multirow}
\usepackage[normalem]{ulem}
\usepackage{soul}
\usepackage{longtable}
\usepackage{mathtools}
\usepackage{hhline}
\usepackage{comment}
\usepackage{appendix}
\usepackage{lastpage,fancyhdr,graphicx}
\usepackage{epstopdf}
\usepackage{setspace} 

\newcolumntype{+}{!{\vrule width 2pt}}
\newlength\savedwidth

\raggedright
\usepackage[aboveskip=1pt,labelfont=bf,labelsep=period,justification=raggedright,singlelinecheck=off]{caption}

\makeatletter
\renewcommand{\@biblabel}[1]{\quad#1.}
\makeatother

\fancyheadoffset[L]{2.25in}
\fancyfootoffset[L]{2.25in}
\newcommand{\frr}{f_r}
\newcommand{\fii}{f_i}
\newcommand{\fdd}{f_d}

\newcommand{\lyt}{\mathrm{lyt}}
\newcommand{\bud}{\mathrm{bud}}

\begin{document}
\vspace*{0.2in}

\begin{flushleft}
{\Large
\textbf{A mathematical model for fitness effects on viral persistence%A mathematical model for fitness effects on viral persistence}%A mechanistic modeling framework links viral fitness to persistence dynamics: experimental evidence from hepatitis C virus} % Please use "sentence case" for title and headings (capitalize only the first word in a title (or heading), the first word in a subtitle (or subheading), and any proper nouns).
}}

Oriol Llopis-Almela\textsuperscript{1,2,*},
J. Tomás Lázaro \textsuperscript{3,1,4},
Antoni Duran\textsuperscript{5},
Celia Perales\textsuperscript{5,6,7},
Esteban Domingo\textsuperscript{5},
Josep Sardanyés\textsuperscript{1}
\\
\bigskip
\textbf{1} Centre de Recerca Matemàtica (CRM). Campus de Bellaterra. Edifici C, 08193 Cerdanyola del Vallès, Barcelona, Spain.
\\
\textbf{2} Departament de Matemàtiques, Universitat Autònoma de Barcelona. Campus de Bellaterra. Edifici C, 08193 Cerdanyola del Vallès, Barcelona, Spain.
\\
\textbf{3} Departament de Matemàtiques. Universitat Politècnica de Catalunya - BarcelonaTech (UPC). Avda. Diagonal 647, 08028 Barcelona, Spain.
\\
\textbf{4} Institute of Mathematics, UPC-BarcelonaTech (IMTech). C. Pau Gargallo 14, 08028 Barcelona, Spain.
\\
\textbf{5} Microbes in Health and Welfare Program, Centro de Biología Molecular Severo Ochoa (Consejo Superior de Investigaciones Científicas, CSIC-Universidad Autónoma de Madrid), Madrid 28049, Spain.
\\
\textbf{6} Department of Clinical Microbiology, Instituto de Investigación Sanitaria-Fundación Jiménez Díaz University Hospital, Universidad Autónoma de Madrid, Madrid 28040, Spain. 
\\
\textbf{7} Centre for Biomedical Network Research on Infectious Diseases (CIBERINFEC), Instituto de Salud Carlos III, Madrid 28029, Spain.

\bigskip

* Corresponding author: Oriol Llopis-Almela (ollopis@crm.cat)

\end{flushleft}

\section*{Abstract}
Persistent viral infections arise from complex interactions between viral replication, host cell responses, and ongoing viral evolution. A general framework linking viral fitness to persistence dynamics is lacking. Here, we develop for the first time a mathematical model of viral persistence that takes into consideration viral fitness variations. Essential to the model is the partition of the classical fitness parameter into three components: replicative, infective, and dispersal fitness. The model was initially inspired by a new experiment on hepatitis C virus (HCV) persistence, established in human hepatoma cells, also reported in this work. This experiment documents two strikingly different viral trajectories depending on the initial replicative fitness of the viral population used to establish persistence. The dynamical model describes the interactions among uninfected cells, infected cells, and infectious virions, and it incorporates, through a continuum, two alternative mechanisms of viral release from cells: budding and lysis. Analysis of the model reveals that viral fitness parameters organise infection outcomes into distinct dynamical regimes. Low replicative and dispersal fitness values lead to viral extinction, whereas high values enable persistence through either stable coexistence or recurrent infection waves. The space of fitness discloses a hierarchy among these parameters, with replicative and dispersal fitness being able to trigger important shifts in the outcome of the infection as opposed to infective fitness. The model identifies trade-offs between replication and dispersal that shape viral production and predicts slow dynamical regimes in which infection may persist despite low detectable viral loads. These dynamical transitions provide candidate mechanisms capable of generating the persistence patterns observed experimentally. Our results establish a computational framework linking multidimensional viral fitness to persistence dynamics and suggest general principles by which evolving RNA viruses transition between extinction (cell curing) and sustained persistence. \\

\bigskip
\textbf{Keywords:} RNA viruses, viral persistence, viral fitness, bifurcations, complex systems, dynamical systems

\bigskip
\textbf{Experimental data and simulation code:} they are available at \url{https://github.com/josepsardanyes/virus_fitness_dynamics_modeling}

\clearpage
\newgeometry{top=0.85in,left=1in,right=1in,footskip=0.75in}
%\linenumbers

\section{Introduction}
Long-term, persistent viral infections have been described in many biological systems. For humans, they pose a continuing medical and public health challenge.  This is amply documented with latency and reactivation of some DNA viruses (with a corresponding disease recrudescence), via integrated DNA or episomic DNA, and changes in gene expression programs (herpes simplex, Epstein Barr, adeno-associated).  Sequels can follow after resolution of acute infections by RNA viruses as diverse as influenza, measles or polio by multifactorial processes that are not well understood, but that may involve persistence of viral genomes in modified forms \cite{Leparc-Goffart1996EvidenceSyndrome, Punsoni2023Post-PolioRevisited, Dutton2025HerpesLifespan, Fu2025MechanismsReactivation, Hirai2025AminoTriggering}. \\

Atypical disease manifestations subsequent to an acute viral disease episode, are currently exemplified by Long COVID [or Long COVID Syndrome (LCS)], as an outcome affecting around 15\% of COVID-19 patients \cite{Davis2023LongRecommendations}. Viruses can persist in their host organisms either asymptomatically, sometimes in quasi-symbiotic relationships \cite{Virgin2009RedefiningInfection, Goic2012LivingViewpoint,Roossinck2015MetagenomicsLifestyles,Dias2025PersistentSurvival, Hull2026RNAOrganisms}, or with pathological manifestations which are different from those observed in the acute infection phase. This meaning of within-cell persistence should be distinguished from the long-term virus survival in nature, either through acute or persistent infections, or both. This second meaning of persistence is the result of trade-offs between viral replication, transmission, pathogenesis, and host fitness \cite{Lidsky2017VariabilityInfections}. While some concepts (i.e., virulence) are relevant to the two meanings of persistence, aspects such as host population density, migration or host-to-host viral transmission play no direct role in persistence within an individual host, a distinct process in itself endowed with remarkable complexity. \\

Several interconnected influences have to converge to achieve survival of cells (or organism) and the resident virus. The latter must somehow curtail the host antiviral response or modify its own genome structure and gene expression program to prevent virus clearance. To the aim of coexistence, DNA and RNA viruses exhibit multiple mechanisms that can be divided in two broad classes, epitomized with the terms \emph{interaction} and \emph{mutation-evasion} \cite{Ahmed1999PersistentInfections, Oldstone2006ViralPredictions, Fan2018BreakingCycle, Sigal2025ThePersistence, Case2025SARS-CoV-2:Immunity}. Most viruses exploit the two classes of mechanisms, albeit to different extents. Complex DNA viruses (i.e., herpes viruses, poxviruses or asfarviruses) are particularly adept in interaction strategies, while the RNA viruses (i.e., influenza virus, hepatitis C virus or the SARS coronaviruses) are particularly adept in mutation-evasion strategies. \\

In the interaction strategy, a balance is reached between continued viral replication and a proper dosage of cellular effectors of innate and adaptive immunity. Immune modulation may consist in the boosting of an immunosuppressive environment [ie. by enhancing IL-10 and TGF-$\beta$ secretion which inhibits inflammatory responses, cytotoxic T lymphocyte (CTL) (including T cell repertoire shrinkage) and natural killer (NK) cell activity], downregulation of the major histocompatibility complex (MHC) class I, disruption of interferon (IFN) and cellular signaling pathways (i.e., JAK / STAT), or alteration of apoptosis (programmed cell death) \cite{Virgin2009RedefiningInfection, Ahmed1999PersistentInfections, Oldstone2006ViralPredictions, Case2025SARS-CoV-2:Immunity}. To contribute to the modulation, viruses may encode homologues of cytokines, chemokines and their receptors.  The immune effector mechanisms may differ depending on the cell types where persistence is established, including particular cases such as immune-privileged sites within an organism \cite{Alcami2003ViralReceptors, Seet2003PoxvirusesEvasion, Oldstone2006ViralPredictions, Virgin2009RedefiningInfection, Eslami2025DecipheringInfection}. There are organ-specific viral reservoirs, with episodic shedding of viral subpopulations \cite{Blois2025SheddingInfection}, suggesting frequent modifications of viral fitness depending on the local environment. To add to the complexity, metabolic factors, in particular modifications of lipid metabolism may also favor lytic versus chronic or latent viral interaction modes (see \cite{Lei2025PINLYP-mediatedInfection} and studies quoted therein). Given the human genome diversity among populations and individuals of the same ethnicity, and the inherent virus trait heterogeneity, the response to a viral infection is variable and unpredictable \cite{Holland1992RNAQuasispecies, Virgin2009RedefiningInfection}. \\

The triggering of autoimmune disease is not alien to host-virus interactions.  Molecular mimicry (or the similarity of protein stretches in viral proteins and host proteins as perceived by the immune system) may result in the induction of pathological self-antibodies, directed to a variety of host proteins, including IFN-1 and nuclear proteins \cite{Oldstone1998MolecularDiseases, Hromic-Jahjefendic2025TheDiseases}. These several effects that result from viral proteins influencing host functions permit continuing virus replication; host pathology, when present, can be viewed as a side effect of interactions that have an ancient origin in life history \cite{Domingo2020VirusImplications, Hull2026RNAOrganisms}. \\

The mutation-evasion strategy ---which is compatible with the interaction strategy--- has also several facets. It can be based on mutations that decrease the genome replication rate, or that limit virus entry into the cell, or that confer resistance to two major components of the immune response: antibodies (Ab) that target and inactivate viral particles, or cytotoxic T-lymphocytes (CTLs) that target and eliminate infected cells. The preference of RNA viruses for mutation-evasion is imposed by the compactness (limited coding capacity) of their RNA genome. Indeed, most RNA virus genomes are in the range of 4,000 to 31,000 nucleotides, while complex pathogenic DNA virus genomes often comprise 130,000 to 360,0000 base pairs, and they encode 80 to 300 proteins, with about one third of them involved in immune modulation. RNA viruses rely on the high mutation rates exhibited by their RNA-dependent RNA polymerases (RdRp) as an adaptive strategy, a trait which is tolerated partly by the limited RNA genome size \cite{Schuster2023VirusLandscapes}. High error rates during replication underlie quasispecies dynamics (with participation of viable and defective versions of viral genomes) and result in diverse and continuously changing populations characterised by complex mutant spectra (also termed mutant clouds or mutant swarms) in each infected individual.  The power of the mutation-evasion mechanism is illustrated by the fact that a comparable role in persistence necessitates the participation of 20 to 100 virus-coded proteins in complex DNA viruses. Genome flexibility confers escape potential to host responses, but individual mutations per se (single amino acid substitutions) in critical viral proteins can modify the infection phenotype from lytic to persistent \cite{Young2019TheSubunit, Hirai2025AminoTriggering}. \\

Not surprisingly, the subset of RNA virus subpopulations that establish persistence is not identical (and often genetically quite different) from the initial virus that caused the acute infection in the same host species. By virtue of these differences, the persistent viruses may replicate in host sites other than those where the precursor acute viruses replicated. Consequently, they may produce symptoms that are unrelated to those caused by the standard form of the initial acute virus \cite{Holland1982RapidGenomes, Gebauer1988RapidCattle, Cattaneo1992MutationsInfections, Holland1992RNAQuasispecies, Taye2024AGenomes, Hirai2025AminoTriggering}. This is again illustrated by the about 200 different symptoms of LCS \cite{Davis2023LongRecommendations, Rong2024PersistenceCOVID-19}. \\

Cell culture models of viral persistence have been developed. They do not capture the complexities and nuances of \emph{in vivo} persistence, but they serve to focus on a subset of the variables that contribute to a joint and prolonged virus and cell survival. Two major modes of cell culture viral persistence were initially distinguished: the virus-cell carrier system and the steady-state persistence \cite{Fenner1976MedicalVirology}. In the carrier state mode, a fraction of uninfected cells is present; these cells are re-infected by virus produced by cell subsets that exhibit, on average, absent or limited cytopathology. In the steady-state mode, no uninfected cells are present. Among many persistent infections in cell culture that have been studied to date, those that display dynamics of cell and virus co-evolution are of particular interest (several examples reviewed in \cite{Domingo2020VirusImplications}). These systems introduced the concept of viral fitness variations, notably fitness dependence on the host cell that coevolved with the virus. As examples, increased viral virulence for the parental host cells was a trait acquired upon long-term persistence in cell culture infections of foot-and-mouth disease virus \cite{delaTorre1988CoevolutionCulture} and hepatitis C virus (HCV) \cite{Zhong2006PersistentAndHost}. Virus-cell coevolution and other features of persistent viral infections are potentially tractable mathematically, although no theoretical models that recapitulate the several facets of persistence in cell culture have been developed. Previous models have approached viral persistence considering modulation of the immune response, virus escape, and virus hiding in the form of viral DNA integration in host DNA, among other aspects \cite{Ganusov2013MathematicalResponses, Lim2014HTLV-IImmunity, Browne2018DynamicsNetwork, Gulbudak2021ACells, Lidsky2017VariabilityInfections}. Mathematical models have also addressed lytic versus lysogenic microbial infections by bacteriophages in the context of microbial ecology \cite{Weitz2019ViralLatency, Gulbudak2019HeterogeneousSystems}. \\

Despite being a relevant parameter for the consequences of virus-host interactions, viral fitness has rarely entered experimental designs or theoretical models of viral persistence. There is a need for a general theoretical framework for intra-host viral persistence that contemplates viral population dynamics and fitness variations as central ingredients. In the present study, we develop a theoretical model of viral and host dynamics, taking as initial experimental reference the cell culture carrier cell systems, with limited cell killing and cell re-infection, of which the steady-state persistence can be considered a particular case (with the absence of uninfected cells). The parameters (and their likely ranges, based on measurements in infections with different viruses), introduced in the model are: the rate of division of uninfected and infected cells, the rate of generation of uninfected cells from infected cells, self-curing rate of infected cells by division, virus burst size and death rate of infected cells, rate of virus release into the external medium, virus fitness and inactivation rate. \\

Importantly, we introduce two novelties which have not been considered in previous treatments: (i) that the viruses we are dealing with are mutant clouds, not defined nucleotide sequences \cite{Sardanyes2024QuasispeciesApplications, Domingo2025AQuasispecies}, and (ii), as a consequence of the former, that viral fitness \cite{Domingo2019ViralInterventions} has continuously changing values (moving target nature) in the course of a persistent infection. To deal with the latter aspect, we have split the classic viral fitness concept \cite{Domingo2019ViralInterventions} into three sub-classes: replicative, dispersive and infective fitness. Once defined, they are placed in distinct parameters of the model. Predictions of the model are confronted with new experimental data on persistence in human hepatoma cells in culture, by comparing two persistently infected cultures that were initiated by two HCV populations that differed by 2.3-fold in initial fitness. We delineate how the model can stimulate new experiments to deepen our understanding of viral persistence, in particular for large RNA viruses such as SARS-CoV-2, whose fitness may be prone to decrease as a consequence of the introduction of escape mutations \cite{Schuster2023VirusLandscapes}.

\section{Materials and methods}
\subsection{Cell lines, viruses and cell passages}
Huh-7 cell line was used to obtain persistently infected cells. These cells, originally derived from a liver tumour of a patient with hepatocellular carcinoma, exhibit stable genetic traits and are readily culturable. They are susceptible to HCV infection~\cite{Nakabayashi1982GrowthMedium.}. 

The initial HCV population used to infect the cell cultures was generated via transcription of the plasmid Jc1FLAG2 (p7-nsGluc2A) followed by cell transfection \cite{Perales2013ResponsePhenotype}. This was amplified in human hepatoma cells to yield HCV p0.  HCV p0 was subsequently used to derive the high-fitness variant HCV p200 in a non-coevolving cellular environment by performing 200 serial passages of the virus shed into the cell culture supernatant in Huh-7.5 reporter cells (fresh cells were used at each new passage, so that no cell passages were involved in the preparation of HCV p200) \cite{Moreno2017InternalEnvironment}. A primary phenotypic difference observed between HCV p0 and HCV p200 was the increased progeny production resulting in a 2.3-fold higher viral fitness of HCV p200 than HCV p0. In this context, viral fitness refers specifically to replicative fitness \cite{Moreno2017InternalEnvironment, Sheldon2014IncreasedVirus}.
%\linelabel{bl}
Notably, HCV p200 also exhibited an enhanced capacity for cell killing, with an increase in virus burst size and the release of virions into the medium. Thus, the HCV p200 was labelled as a mainly lytic virus. Conversely, cell killing was not observed with HCV p0, and virion release took place through other mechanisms, such as budding from the cell membrane. In what follows, we denote as budding any mechanism of the virus to exit the cell other than cell lysis. \\

Huh-7 cells were cultured at 37$^{\circ}$C with 5\% CO2 in Dulbecco's modified Eagle's medium (DMEM), supplemented with 10\% Fetal Bovine Serum \cite{Moreno2017InternalEnvironment}. Each cell line was analysed in three independent replicates for both HCV p0 and HCV p200 conditions. Initially, the cell monolayer in each replicate was infected at a multiplicity of infection (MOI) of 0.03 TCID50/cell (where 50\% Tissue Culture Infectious Dose represents the amount of virus required to infect 50\% of the cell monolayer and the MOI stands for the number of infectious virions per cell). Following a 5-hour virus adsorption period and subsequent incubation ---during which cells became infected and massive cell lysis occurred in HCV p200 infection--- the inoculum was removed, and 2 mL of fresh medium was added to the monolayer. This step was critical for HCV p200, as only a few infected cells remained attached, and detachment from the monolayer was probable. In contrast, the probability of cell monolayer detachment was low for HCV p0 due to its limited lytic behaviour. \\

\begin{figure}[ht!]
    \centering
    \includegraphics[width = \linewidth]{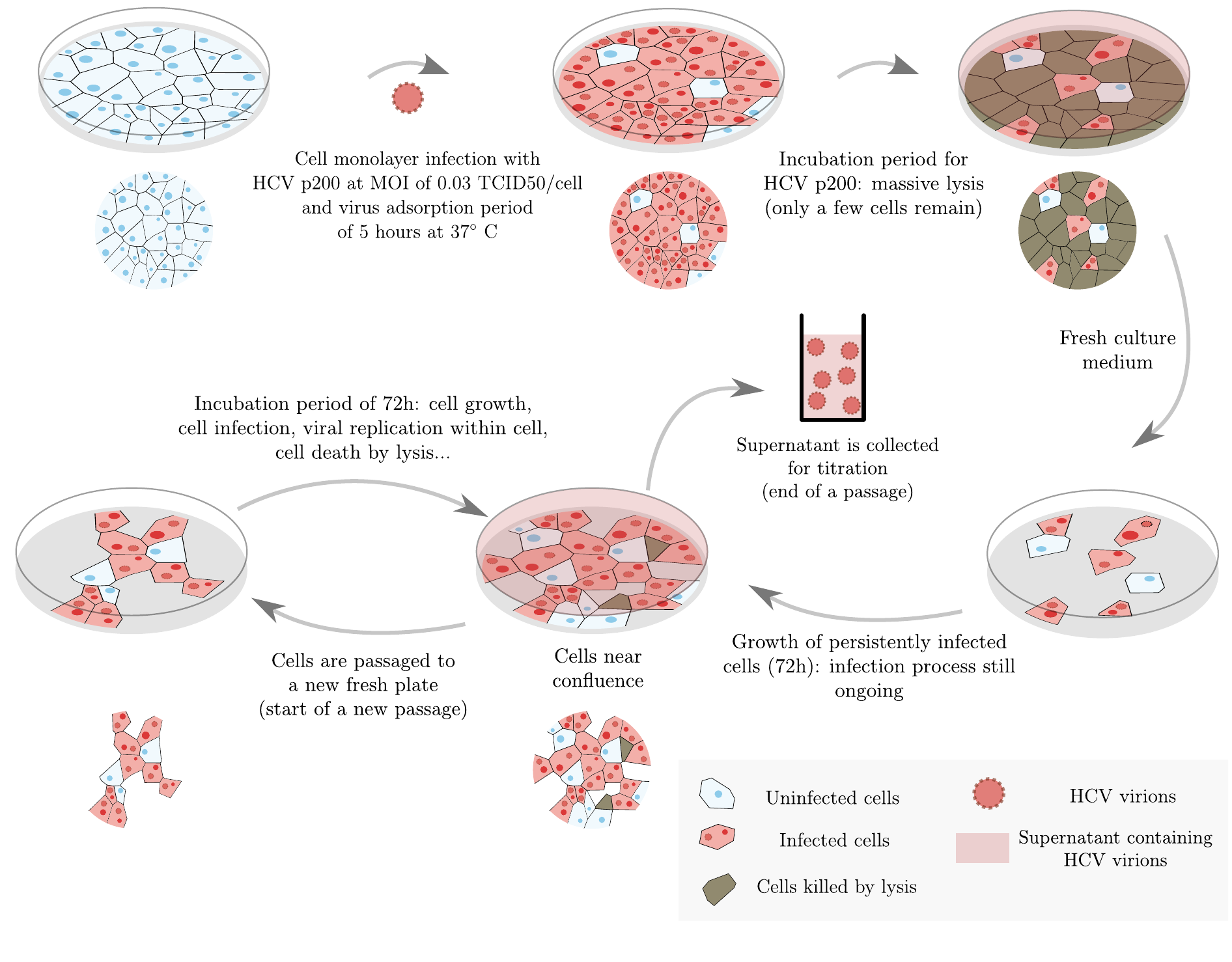}
    \caption{\textbf{Schematic representation of the experimental procedure.} Cell culture steps to establish persistent infections with HCV in Huh-7 cells. The experiment corresponds to infection by HCV p200, which resulted in massive cell killing. The major steps summarised from top to bottom, and linked by curved arrows, are: cells are infected with virus (particle in pink colour, not drawn to scale), massive cell killing results in a few surviving cells that establish the carrier state; cells can grow to near-confluence, and they can be passaged, frozen, thawed and re-cultured while maintaining the capacity to shed infectious virus into the cell culture medium. Biological drawings differed in the infections by HCV p0 because the initial cell killing was more limited. Events summarised in the scheme provide the parameters incorporated in our mathematical model.}
    % \caption{Protocol for generating persistently infected cells via serial passaging of an initial monolayer infected with the HCV strain p200 (obtained from serial passaging in~\cite{Moreno2017InternalEnvironment}). Biological dynamics differ when the monolayer is initially infected with the HCV strain p0: cell lysis does not occur, allowing the culture to reach confluence rapidly prior to the first passage.}
    \label{cell_infection}
\end{figure}

Cells were then incubated for 72 hours, a period during which cell growth, viral genome replication, and virion assembly occurred. Throughout this time, cells grew until reaching complete or nearly complete confluence. Subsequently, cells were passaged into fresh medium, and the infectious cell culture supernatant was collected for titration of infectivity, extracellular viral RNA quantification, and calculation of specific infectivity (the ratio between infectivity and amount of viral RNA; not all viral RNA is infectious due to the presence of various forms of viral RNA with lethal lesions). Thus, at the start of each passage, no free virions were present in the supernatant. The objective of this passaging protocol was to assess the virus's capacity to persist in the cells. The protocol for HCV p200 is illustrated in Fig~\ref{cell_infection}. 

\subsection{Mathematical model}
Our modelling approach builds upon the experimental studies designed to assess the impact of viral fitness on the establishment and evolution of HCV persistence in cell cultures, described in the previous section. Let $c_p(t)$, $y_p(t)$ and $V_p(t)$ denote the population of uninfected cells, infected cells, and infectious virions in the supernatant of the cell culture at time $t$ of passage $p$, respectively. Each passage $p$ begins at time $t_0 = 0$ hours (h) and ends at time $t_f = 72$ h, the time elapsed between two successive cell passages. Since the variables stand for populations, we restrict them to the biologically meaningful space
\begin{equation}
    \mathcal{V} = \left\{(c_p(t),y_p(t),V_p(t)) \in \mathbb{R}^3 \quad | \quad c_p(t) \geq 0, y_p(t) \geq 0, V_p(t) \geq 0\right\},
\end{equation}
Each of the main abilities that allow viruses to complete a replication cycle successfully is described by a different fitness term, whose distinction is relevant to our model:
\begin{itemize}
    \item the replicative fitness $\frr$ describes the ability of the virus to take advantage of the cell's resources to replicate the viral genome,
    \item the infective fitness $\fii$ describes the aptitude of the virus to infect a cell, and
    \item the dispersal fitness $\fdd$ measures the success in becoming a stable particle and leaving the cell via budding or by causing cell lysis.    
\end{itemize}

\begin{figure}[ht!]
    \centering
    \includegraphics[width=\linewidth]{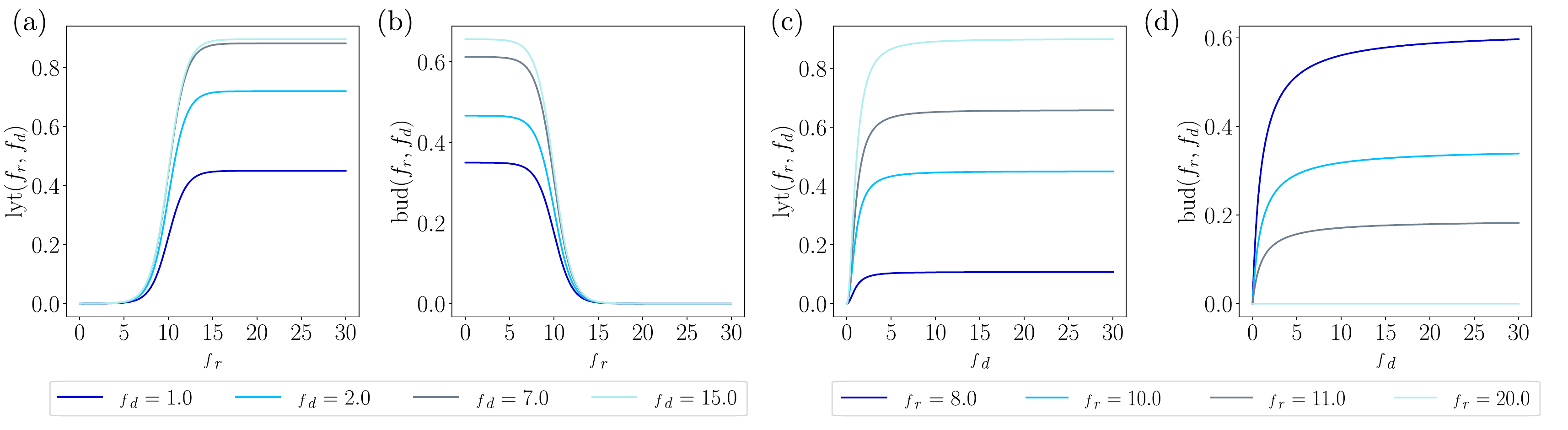}
    \caption{\textbf{Lysis and budding weight functions.} (a, b) Weight functions of the lytic process $\lyt(\frr,\fdd)$ and the budding process $\bud(\frr,\fdd)$ as a function of the replicative fitness $\frr$ for different values of $\fdd$. (c, d) Analogous to (a) and (b), as a function of $\fdd$ for different values of $\frr$. The values used for the parameters in these functions are: $b_{\ell} = 0.9$, $b_b = 0.7$, $c_1 = e^{10}$ and $\gamma = 10^{-6}$.}
    \label{lytbudfunc}
\end{figure}

Each of these fitness values is inherently variable due to the high mutation rates that characterise RNA viruses \cite{Sanjuan2010ViralRates}. In the context of partitioned fitness values, the dynamics within the cell culture at passage $p$ can be described by the following system of ordinary differential equations, which represent the time evolution of the populations:
\begin{align}
    \frac{dc_p}{dt} &= \left(\alpha_c c_p + \alpha_y \kappa y_p \right) \left(1 - \frac{c_p + y_p}{K} \right) - \inf(\fii) V_p c_p, \label{dcdt}\\
    \frac{dy_p}{dt} &= \alpha_y (1 - \kappa) y_p  \left(1 - \frac{c_p + y_p}{K}\right) + \inf(\fii) V_p c_p - (\delta_{\ell} \, \lyt(\frr,\fdd) + \delta_d) y_p, \label{dydt}\\
    \frac{dV_p}{dt} &= (\delta_{\ell} \, \lyt(\frr,\fdd) + \bud(\frr,\fdd)) \frr y_p - \inf(\fii) c_p V_p - \varepsilon V_p, \label{dVdt}
\end{align}
where functions $\inf(\fii)$, $\lyt(\frr,\fdd)$ and $\bud(\frr,\fdd)$ stand for the strength of the cell infection process, the death of the cell by lysis, and the release of virions to the supernatant via budding, respectively, as functions of the three fitness types. They are described by the following expressions:
\begin{equation}\label{inf_func}
    \inf(\fii) = \gamma \frac{\fii}{1 + \fii},
\end{equation}
\begin{equation}\label{lyt_func}
    \lyt(\frr,\fdd) = \underbrace{\frac{1}{1 + c_1 \exp(-\frr)}}_{\substack{\textrm{lytic effect will be} \\ \textrm{predominant for viruses} \\ \textrm{with high replicative} \\ \textrm{fitness}}} \cdot \underbrace{\beta_{\ell}(\fdd)}_{\substack{\textrm{amplification of} \\ \textrm{the lytic effect} \\ \textrm{due to the dispersal} \\ \textrm{fitness}}}; \quad \beta_{\ell}(\fdd) = \underbrace{b_{\ell}}_{\substack{\textrm{burst} \\ \textrm{size}}} \frac{\fdd^2}{1 + \fdd^2},
\end{equation}
\begin{equation}\label{bud_func}
    \bud(\frr,\fdd) = \underbrace{\frac{1}{1 + c_1^{-1} \exp(\frr)}}_{\substack{\textrm{budding will be} \\ \textrm{predominant for viruses} \\ \textrm{with low replicative} \\ \textrm{fitness}}} \cdot \underbrace{\beta_{b}(\fdd)}_{\substack{\textrm{amplification of} \\ \textrm{the budding effect} \\ \textrm{due to the dispersal} \\ \textrm{fitness}}}; \quad \beta_{b}(\fdd) = \underbrace{b_{b}}_{\substack{\textrm{maximum number of} \\ \textrm{virions released} \\ \textrm{by budding}}} \frac{\fdd}{1 + \fdd},
\end{equation}
where we assume that all of the functions saturate at a maximum value when the different fitness values increase, and that the lytic and the budding processes are dominant in different ranges of $\frr$. Based on the experimental results obtained with different HCV populations \cite{Moreno2017InternalEnvironment}, viruses with low $\frr$ are predominantly non-lytic, while those with high replicative fitness are mainly lytic (Fig~\ref{lytbudfunc}a). Eqs~\eqref{lyt_func}-\eqref{bud_func} describe the dominance of each exit strategy as a function of $\frr$, accounting for the non-exclusivity of each of them in a viral population. That is, both exit strategies coexist for all values of $\frr$ but with different weights. \\
%\textcolor{blue}{CELIA: Entiendo que queda claro en algún sitio del documento que los procesos de budding y lisis no son mutuamente excluyentes si no que pueden coexistir...y que predomina uno u otro según según el fitness....?} \\
%\textcolor{blue}{ORIOL: Esto se menciona, por una parte, en la línea~\ref{bl}. Por otra parte, tres líneas más arriba hablamos de ``... the lytic and the budding processes are dominant in different ranges of $\frr$''. También, las funciones lyt y bud nunca valen cero tal como están descritas, hecho que da lugar a que puedan coexsitir ambos valores. Estos son los tres puntos en que aparecen referencias a lo que comentas. ¿Crees que deberíamos reforzarlo más?} \\

The amount of virions released into the cell culture medium depends upon the number of particles assembled within the cell, and their ability to exit the cell. Thus, $\fdd$ modulates the extent of the exit process with a maximum burst size $b_{\ell}$ (Fig~\ref{lytbudfunc}c). Analogously, non-lytic viruses are mainly released to the supernatant by encapsidation through the cell membrane for low $\frr$ (Fig~\ref{lytbudfunc}b), and the number of virions released will depend upon the dispersal ability of the virus. This effect is modulated by the function $\beta_{b}(\fdd)$ (Fig~\ref{lytbudfunc}d), which is assumed to grow as $\fdd/(1+\fdd)$, shifting from very small values of $\fdd$ and saturating at a maximum budding release $b_b$. On the other hand, the amplification function of the lytic effect $\beta_{\ell}$, which requires a minimal value of $\frr$ to be manifested, shows a more explosive effect by growing faster than $\beta_b$ at low $\fdd$ values. It saturates at a maximum release $b_\ell$, and in this case, a sigmoidal-like function $\fdd^2/(1+\fdd^2)$ has been chosen. These functions contrast the smooth nature of budding with the bursting nature of lysis. Thus, we emphasise the key role of $\frr$ and $\fdd$ in determining the lytic behaviour of the virus, according to the experimental evidence \cite{Moreno2017InternalEnvironment}.

Eqs~\eqref{dcdt}-\eqref{dVdt} describe the cellular processes of division within the cell culture and competition for available resources, which are limited both by the space and the medium: cells can grow until reaching confluence in the monolayer, and the environment in which they evolve must be fresh enough to ensure cellular division. Infected cells are assumed to degrade at a faster rate than uninfected cells, whose death rate we consider negligible \cite{Kofahi2016HepatitisCells, Deng2008HepatitisPathway}.

Each process has crucial parameters, shown in Table~\ref{param_table}. It includes the corresponding magnitude of each parameter, the unit of measure and the value that has been used to perform numerical simulations according to the estimations computed for HCV \cite{Moreno2017InternalEnvironment}.

\begin{table}[!ht]
    \centering
    \begin{threeparttable}
    \begin{tabular}{c|c|c|c|c}
        \textbf{Parameter} & \textbf{Description} & \textbf{Magnitude} & \textbf{Units} & \textbf{Value} \\
        \hline
        $\alpha_c$ & Division rate of uninfected cells & time$^{-1}$ & hours & 0.023 \\ 
        $\alpha_y$ & Division rate of infected cells & time$^{-1}$ & hours & 0.012 \\ 
        $K$ & Plaque's carrying capacity & Dimensionless & individuals & $10^6$\\ 
        \multirow{2}{*}{$\kappa$} & \multirow{2}{8cm}{\centering Fraction of infected cells that, by division, gives rise to uninfected cells}  & \multirow{2}{*}{Dimensionless} & \multirow{2}{*}{-} & \multirow{2}{*}{$10^{-2}$}\\ 
        & & & & \\
        $\gamma$ & Maximum virus' infection rate & time$^{-1}$ & hours & $10^{-6}$\\ 
        $\delta_{\ell}$ & Death rate due to lysis of infected cells & time$^{-1}$ & hours & 0.025\\ 
        $\delta_{d}$ & Degradation rate of infected cells & time$^{-1}$ & hours & $10^{-1}$ \\ 
        $b_{\ell}$ & Maximum fraction of virions burst (burst size) & Dimensionless & - & 0.9 \\
        $b_{b}$ & Maximum fraction of virions released by budding & Dimensionless & - & 0.7 \\
        $\fii$ & Infective fitness & Dimensionless  & - & - \\ 
        $\frr$ & Replicative fitness & Dimensionless & - & - \\ 
        $\fdd$ & Dispersal fitness & Dimensionless & - & -\\ 
        $\varepsilon$ & Virus' inactivation rate & time$^{-1}$ & hours & 0.0123\\
        $c_1$ & Sigmoid constant & Dimensionless & - & $e^{10}$\\ \hline
    \end{tabular}
    \begin{flushleft}
    \end{flushleft}
    \end{threeparttable}
    \caption{\textbf{Parameters of the model.} Parameters of the model described by Eqs~\eqref{dcdt}-\eqref{dVdt} and~\eqref{inf_func}-\eqref{bud_func}, their magnitude, unit of measurement and value used in the simulations. Appendix~\ref{AppA} contains the details about the values chosen for the HCV case.}
    \label{param_table}
\end{table}

All parameters in Table~\ref{param_table} are assumed to be strictly positive since they are either rates of growth, fractions or quantities referring to populations. Fitness values are assumed to be strictly positive: any of the fitness with zero value would imply a virus unable to perform the associated action (replicate, infect or disperse). Increasing the values of any of the fitness parameters results in the enhancement of the ability associated with that fitness. The three-dimensional space of fitness is defined as:
\begin{equation}
    \mathcal{F} = \left\{ (\frr,\fii,\fdd) \in \mathbb{R}^{3} \quad | \quad \frr, \fii, \fdd \geq 0 \right\}.
\end{equation}
We stress that the range of values used for $\frr$, $\fii$ and $\fdd$ is arbitrary, given the theoretical nature of these parameters: although $\frr$ is actually measured experimentally, its value is always established relative to the fitness value of a reference viral population \cite{Domingo2019ViralInterventions}.  \\

The dynamics from one passage to the next describe the transfer of cells to a new culture dish with fresh medium. Thus, at the beginning of each passage, the viral titer is given by $V(t_0) = 0$. This condition requires a stable, nearly confluent monolayer that encompasses both infected and uninfected cells to proceed to the next passages and ensure sustained persistence. Assuming that, experimentally, a representative portion of the cells is passaged to a new culture dish, the following equations set the initial conditions of the cell culture, i.e., at time $t_0$ of passage $p$ for $p = 1, 2 \dots$:
\begin{align}
    c_p(t_0) &= \frac{c_{p-1} (t_f)}{c_{p-1}(t_f) + y_{p-1} (t_f)} N, \label{c:passage} \\
    y_p(t_0) &= \frac{y_{p-1}(t_f)}{c_{p-1}(t_f) + y_{p-1}(t_f)} N , \label{y:passage} \\
    V_p (t_0) &= 0, \label{V:passage}
\end{align}
% and $\epsilon \in \mathcal{U} (10^{-a}, 10^{a})$ with $a = 1,2 \dots$ is a continuous uniform variable introducing possible fluctuations in the proportion of cells passaged due to experimental noise
where $N$ is the total number of cells passaged. We define $c_0(t_0) > 0$, $y_0(t_0) = 0$, and $V_0(t_0) > 0$ since, to start the initial passage, we assume a monolayer of uninfected cells that is infected at a specific MOI. This recurrent inter-passage scheme can be reduced, for its dynamical analysis, to the study of the dynamics within a passage, with varying initial conditions of the form $c_p(t_0) > 0$, $y_p(t_0) > 0$, and $V_p(t_0) = 0$.

\subsection{Numerical methods and simulation remarks}
\textbf{Reformulation of the governing equations for numerical purposes.} We consider the scaled variables $c=\tilde{c} K$, $y=\tilde{y}K$, $V=\tilde{V}K$ to avoid handling numerical values in the order of $K$, which would lead to the propagation of numerical errors. The equations in these variables read as:
\begin{align}
    \dot{\tilde{c}} &= \left( \alpha_c \tilde{c} + \alpha_y \kappa \tilde{y} \right) (1-\tilde{c}-\tilde{y}) - \inf(\fii) K \tilde{V}\tilde{c}, \label{dcdt_n} \\
    \dot{\tilde{y}} &= \alpha_y (1-\kappa) \tilde{y} (1-\tilde{c}-\tilde{y}) + \inf(\fii) K \tilde{V} \tilde{c} - (\delta_{\ell} \, \lyt(\frr,\fdd) + \delta_{d}) \tilde{y}, \label{dydt_n} \\
    \dot{\tilde{V}} &= (\delta_{\ell} \, \lyt(\frr,\fdd) + \bud(\frr,\fdd)) \frr \tilde{y} - \inf (\fii) K \tilde{c}\tilde{V} - \varepsilon \tilde{V}. \label{dVdt_n}
\end{align}
Given the nature of the intrinsic biological processes, the duration of the viral replication cycle and that of cell division differ by orders of magnitude \cite{Chen2022ModelingProtection, Zeuzem2001THEINFECTION, Sender2021TheVirions}. Eqs~\eqref{dcdt_n}-\eqref{dVdt_n} deal with a single time variable that we assume to be the viral time scale. We then rescale the evolution equations of the cells as:  
\begin{align}
    \sigma\dot{\tilde{c}} &= \left( \alpha_c \tilde{c} + \alpha_y \kappa \tilde{y} \right) (1-\tilde{c}-\tilde{y}) - \inf(\fii) K \tilde{V}\tilde{c}, \label{dcdt_n_sigma} \\
    \sigma\dot{\tilde{y}} &= \alpha_y (1-\kappa) \tilde{y} (1-\tilde{c}-\tilde{y}) + \inf(\fii) K \tilde{V} \tilde{c} - (\delta_{\ell} \, \lyt(\frr,\fdd) + \delta_{d}) \tilde{y}, \label{dydt_n_sigma} \\
    \dot{\tilde{V}} &= (\delta_{\ell} \, \lyt(\frr,\fdd) + \bud(\frr,\fdd)) \frr \tilde{y} - \inf (\fii) K \tilde{c}\tilde{V} - \varepsilon \tilde{V},\label{dVdt_n_sigma}
\end{align}
with $0 < \sigma < 1$. This parameter, which we set to $\sigma = 10^{-1}$, speeds up cellular processes and avoids numerically handling different orders of magnitude among variables. By accelerating the dynamics of the cell, more division cycles occur per replication cycle of the virus so that changes in both populations are observable in the same time span.

\textbf{Numerical method for ODE integration.} We used the Taylor method as developed in~\cite{Jorba2005AMethods} to numerically integrate Eqs~\eqref{dcdt_n_sigma}-\eqref{dVdt_n_sigma}. The order of the Taylor series was set at 60, and quadruple precision was used in the numerical integration. Unless stated otherwise, Table~\ref{param_table} lists the parameter values used for numerical integration.

\textbf{Numerical computation of limit cycles.} Numerically, we computed limit cycles by finding fixed points of the Poincaré map associated with our system using Poincaré sections of type $V = k \in \mathbb{R}^+$. The numerical implementation of the method is based on Section 5 in \cite{CarlesSimo1990OnManifolds}.

\textbf{Simulation of multiple passages.} The evolution of the infection is simulated by integrating Eqs~\eqref{dcdt_n_sigma}-\eqref{dVdt_n_sigma} during 72 hours, with the initial conditions given by Eqs~\eqref{c:passage}--\eqref{V:passage}. At each passage, we change the fitness parameters to simulate the effects of mutations on viral fitness. We assume that mutations on the viral population accumulated during each passage exert an average effect on the fitness parameters of the entire population. Specifically, we consider that infected cells transferred from passage $p$ culture to passage $p+1$ culture carry a viral population with altered fitness values ($\frr$, $\fdd$, and $\fii$) with respect to those of the preceding passage. Thus, our approach discretises the continuous evolution of fitness into a series of averaged changes when cells are transferred.

The theoretical nature of the fitness parameters introduced in this work and the difficulty to obtain quantitative data beyond the replicative capacity of a viral population hinder the development of a mechanistic description of how fitness parameters evolve across passages. For this reason, we use sampled values of $\frr$, $\fdd$ and $\fii$ at the beginning of each passage in each of the replicates for the numerical integration.

% Hereafter, unless otherwise stated, all the results and simulations presented have been obtained by numerical integration of Equations~\eqref{dcdt_n_sigma}-\eqref{dVdt_n_sigma} using the Taylor method as developed in~\cite{Jorba2005AMethods}. The values of the parameters are those in Table~\ref{param_table}. Moreover, all the analysis carried out in this section focuses on the dynamics of a single passage, that is, on the dynamics generated by equations~\eqref{dcdt_n}-\eqref{dVdt_n} for a long time $t$. Experimentally, this scenario represents a cell culture infected at a specific MOI and left to evolve for a time window of duration $t$, with no passaging of cells or viruses in the supernatant.  \\

% Results and Discussion can be combined.
\section{Results}
This section is organised as follows: first, we describe the experimental data obtained from the persistently infected cell cultures for two different HCV populations that differ in their replicative fitness. Then, we outline the main results obtained from the model. The analysis of the model focuses on the three viral fitness parameters and their role in sustaining persistent infections. First, we present the possible outcomes of a long-lasting single-passage infection as a function of the three viral fitness parameters. Then, we highlight the results of halting the dynamics of the infection at 72 hours and performing multiple passages. Overall, we aim to assess the role of each viral trait in establishing persistent infections and, thereby, determine how fitness changes due to mutation accumulation shape the evolution of an infection according to our model.

\subsection{Loss of persistence in experimental cell cultures}
\begin{figure}[ht]
    \centering
    \includegraphics[width=\linewidth]{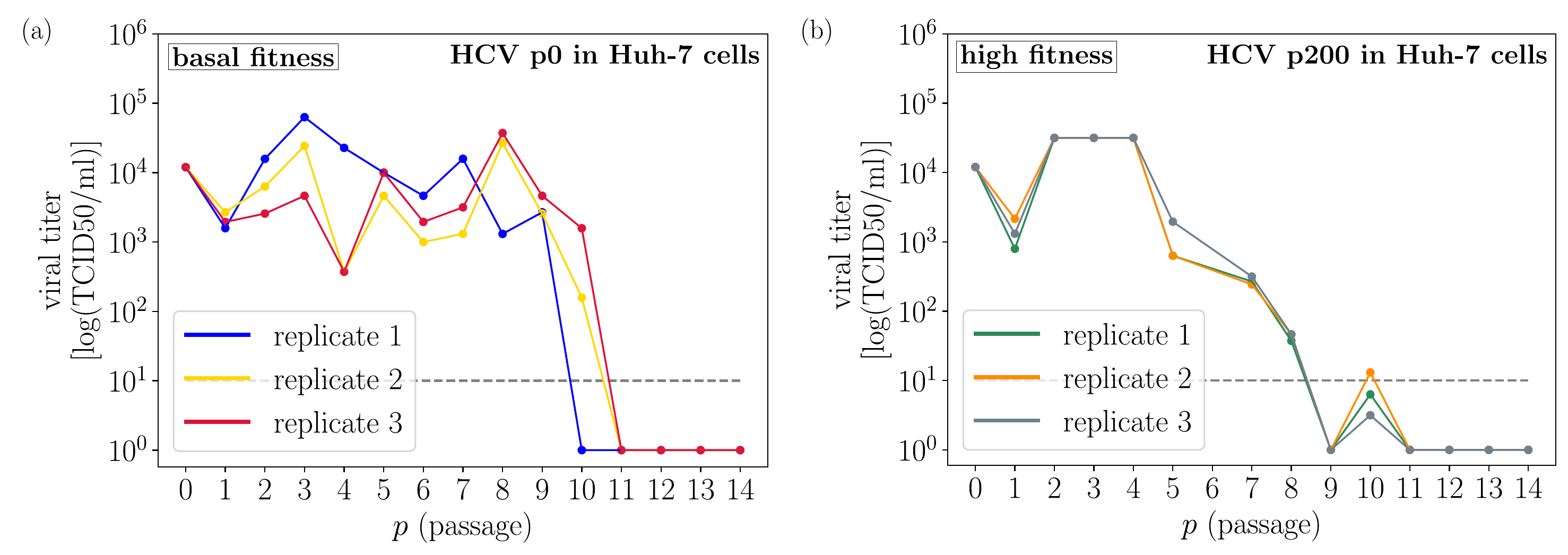}
    \caption{\textbf{Viral titers for each passage in two HCV populations.} Viral titers measured at the end of each passage for (a) HCV p0 and (b) HCV p200 in Huh-7 cells. The data represent three independent replicates for each viral population. The dashed line denotes the experimental limit of detection of viral infectivity.}
    \label{fig:exp:data}
\end{figure}

Cell infection and cell passage were conducted in Huh-7 cells, and three replicates were cultured for each HCV population. Virus titration was performed at the end of each passage, and viral titers in the supernatant for HCV p0 and HCV p200 populations across all passages are presented in Fig~\ref{fig:exp:data}a and Fig~\ref{fig:exp:data}b, respectively. 

A distinct pattern emerged and was shared by the three replicates of HCV p0: non-zero viral titers fluctuated for at least nine passages before decaying abruptly below the limit of detection of infectivity. The HCV p200 infection replicates showed even more consistent titers in the course of cell passage. Titers exhibited a smoother, progressive decline prior to a decrease below the limit of detection, while persistence was sustained in the first four passages. Therefore, depending on the initial replicative fitness of the HCV population, the viral titer decay consistently follows trajectories of different smoothness.

A critical difference between HCV p0 and HCV p200 regarding virus-host cell interaction is that HCV p0 exhibits limited cell killing and, therefore, most virions are released into the cell culture by other mechanisms, such as budding from the cell membrane. In contrast, HCV p200 exhibits a remarkable capacity to lyse cells, resulting in virus shedding into the cell culture supernatant \cite{Perales2013ResponsePhenotype, Moreno2017InternalEnvironment}. In both persistence regimes, when the virus titer reached levels below the experimental limit of detection, the viability of the host cells remained intact. Based on these experimental results with two HCV populations that display different replicative fitness and cell killing capacity, the mechanistic mathematical model described in the former section aims to elucidate the dynamics that drive viral populations towards virus persistence, followed by clearance and loss of the persistent state (at cell passages 9 to 11 in the experiment of Fig~\ref{fig:exp:data}). Among other features, the model captures the observed fluctuations and identifies the viral mechanism underlying the loss of persistence in both HCV populations.

\subsection{Replicative fitness threshold for viral persistence in one-passage dynamics}
By assuming constant dispersal and infective fitness, $f_d = \fii = 1.0$, we focus on the dynamics of a single passage for a long time $t$. Experimentally, this scenario represents a cell culture infected at a specific multiplicity of infection (MOI) and left to evolve for a time window of duration $t$, with no cell or virus passaging. 

Four states of meaningful equilibria, corresponding to four equilibrium points of Eqs~\eqref{dcdt_n_sigma}-\eqref{dVdt_n_sigma}, arise for positive values of $\frr$. The analytical derivation of the meaningful equilibrium points is detailed in the subsequent section. 

\subsubsection{Analytical derivation of equilibrium points as a function of the replicative fitness}\label{ep_deduction_fr}
For the sake of simplicity, we remove the tilde notation from Eqs.~\eqref{dcdt_n}-\eqref{dVdt_n} and we rewrite these equations as:
\begin{align}
    \dot{c} &= \left( \alpha_c c + \alpha_y \kappa y \right) (1-c-y) - \inf(\fii) K Vc, \label{dcdt_nn} \\
    \dot{y} &= \alpha_y (1-\kappa) y (1-c-y) + \inf(\fii) K V c - \Delta_1 y, \label{dydt_nn} \\
    \dot{V} &= \Delta_2 \frr y - \inf (\fii) K cV - \varepsilon V,\label{dVdt_nn}
\end{align}
where
\begin{equation}
    \Delta_1 = \delta_\ell \, \lyt(\frr,\fdd) + \delta_d, \qquad \Delta_2 = (\delta_\ell \, \lyt(\frr,\fdd) + \bud(\frr,\fdd)) \frr.
\end{equation}
By adding equation~\eqref{dcdt_nn} equal to 0 to~\eqref{dydt_nn} equal to 0 and changing the sign of the resulting expression, we get the following equation corresponding to a 2nd-degree polynomial in terms of the variable $y$:
\begin{equation}\label{y_ep_pol}
    \alpha_y y^2 + [c (\alpha_c + \alpha_y) - \alpha_y + \Delta_1] y - \alpha_c c (1 - c) = 0, 
\end{equation}
which, by Descartes' rule of signs, has a single positive and real root: $\alpha_y$ is always positive and also $\alpha_c c (1-c)$, since variables are normalised to 1 by the carrying capacity $K$. Equation~\eqref{y_ep_pol} has the following solution:
\begin{equation}\label{y_eq}
    \psi_{\pm}(c) \coloneqq y_{\pm}(c) = \frac{- c (\alpha_c + \alpha_y) + \alpha_y - \Delta_1 \pm \sqrt{[c (\alpha_c + \alpha_y) - \alpha_y + \Delta_1]^2 + 4 \alpha_c\alpha_y c (1-c)}}{2\alpha_y}.
\end{equation}
From the zero equation of~\eqref{dVdt_nn} and using~\eqref{y_eq}, we obtain:
\begin{equation}\label{V_eq}
    \varphi_{\pm}(c) \coloneqq V_{\pm}(c) = \frac{\Delta_2 \, \psi_{\pm}(c)}{\varepsilon + \inf(\fii) c K}.
\end{equation}
Substituting~\eqref{y_eq} and~\eqref{V_eq} in the zero equation of~\eqref{dcdt_nn}, we obtain the expression that $c$ must satisfy at the equilibrium:
\begin{equation}\label{c_eq}
    (\alpha_c c + \alpha_y \kappa \psi_{\pm}(c)) (1 - c - \psi_{\pm}(c)) - \inf(\fii) c K \varphi_{\pm}(c) = 0.
\end{equation}
Then, points of type $(c^*, y_{\pm}(c^*), V_{\pm}(c^*))$, where $c^* \geq 0$ is a solution of equation~\eqref{c_eq} and $y_{\pm}(c^*) \geq 0$ and $V_{\pm}(c^*) \geq 0$ are computed according to equations~\eqref{y_eq} and~\eqref{V_eq}, give rise to equilibrium states in the non-negative orthant of $\mathbb{R}^3$. \\

The case $c=0$ satisfies equation~\eqref{c_eq} only for $\psi_+(0) = 0$. Equation~\eqref{V_eq} reads as $\varphi_+(0) = 0$, which gives rise to the trivial equilibrium point $Q_0 = (c_0, y_0, V_0) = (0,0,0)$. The case $c=1$ satisfies~\eqref{c_eq} only for, again, $\psi_+(1) = 0$, and not for the counter-case $\psi_-(1)$. Equation~\eqref{V_eq} reads as $\varphi_+(1) = 0$. These values give rise to $Q_1 = (c_1, y_1, V_1) = (1,0,0)$ equilibrium point. \\

Taking equation~\eqref{c_eq} as a function of $\frr$ with $\psi_+(c)$ in equation~\eqref{y_eq} (and, hence, $\varphi_+(c)$ in equation~\eqref{V_eq}), the numerical analysis reveals a maximum of two positive roots for increasing values of the parameter. Thus, two biologically meaningful coexistence equilibrium points arise: $Q_2$ and $Q_3$.

\subsubsection{Bifurcation analysis}
From the previous analysis, our system has two trivial steady states given by $Q_0 = (0,0,0)$ and $Q_1 = (1,0,0)$: the first corresponds to the complete extinction of both cells and the virus, and the second to virus clearance with no cell killing. From the stability viewpoint, the origin $Q_0$ is always a saddle-type unstable equilibrium. It is straightforward to state that:
\begin{itemize}
    \item The axis $\{y = V = 0\}$ is invariant for the dynamics for any value of the fitness. The dynamics are governed by the classical logistic growth of cells under limited resources, given by $\dot{c} = \alpha_c\, c\, (1 - c/K)$. Linearisation around $Q_0$ ($c=0$) gives rise to a linear unstable line with eigenvalue $\lambda_1 = \alpha_c > 0$.
    \item Analogously, the absence of cells $\{c = y = 0\}$ is invariant under the dynamics, which are governed by the viral degradation in the absence of cells given by $\dot{c} = \dot{y} = 0$ and $\dot{V} = -\varepsilon V$. This situation gives rise to a stable (attracting) invariant line to $Q_0$, with associated eigenvalue $\lambda_2 = -\varepsilon < 0$.
    \item The third eigenvalue of the Jacobian matrix of the system in $Q_0$ is real and negative in the range of $\frr$ considered. Its cumbersome expression requires an algebraic manipulator to obtain its analytical expression.
\end{itemize}
Then, the infection can not evolve towards a total loss of the populations since, in the absence of infectious virus, cells are not degraded. 

\begin{figure}[ht!]
    \centering
    \includegraphics[width = \linewidth]{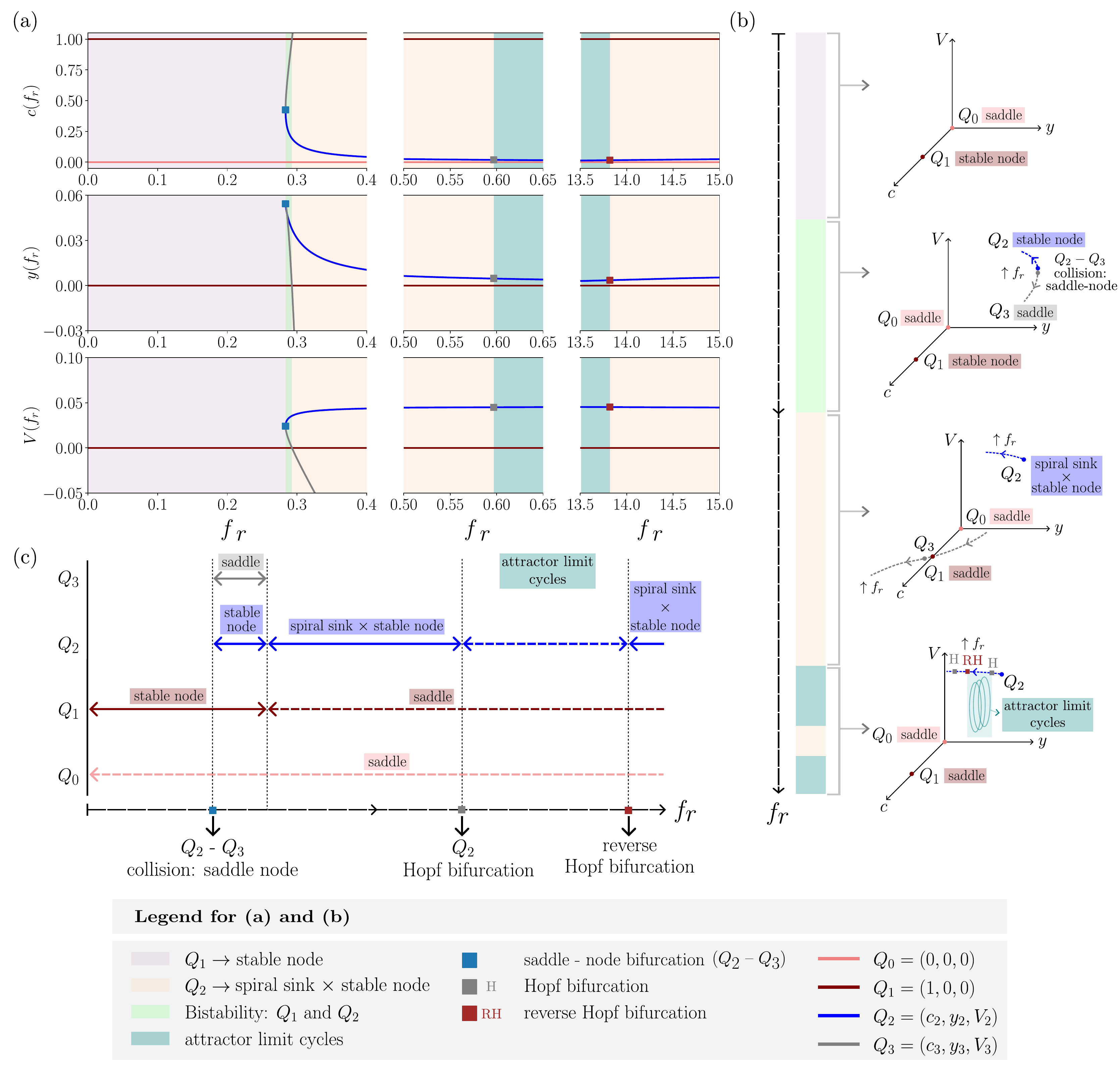}
    \caption{\textbf{Changes in the equilibrium points and in their stability as a function of $\frr$.} (a) Coordinates $c$, $y$, and $V$ (top, middle, and bottom panels, respectively) versus $\frr$ for the biologically feasible equilibrium points of the system described by Eqs~\eqref{dcdt_n}--\eqref{dVdt_n}. Bifurcation points are indicated by squares, and stability regions are shaded in the colours shown in the legend. (b) Schematic diagram of the qualitative dynamics of the system as $\frr$ increases. Dashed arrows denote the direction of increasing $\frr$. (c) Schematic and qualitative display of the local stability of the meaningful equilibrium points as $\frr$ increases. Solid lines indicate stability, while dashed lines indicate instability.}
    \label{fr_ep_coordinates}
\end{figure}

The equilibrium point $Q_1$ lies on the invariant line $\{y = V =0\}$ and it is attained when the carrying capacity $K$ is achieved. Its local stability varies with $\frr$: it is stable for low values of $\frr$ (light violet region in Fig~\ref{fr_ep_coordinates}a and b), and it becomes unstable at a specific value of $\frr \approx 0.295$. That is, viral inactivation predominates over viral replication for low values of $\frr$, driving the infection to viral clearance and cells to reach confluence. Above a lower boundary of $\frr$, the viral population survives, and the infection can be sustained.

Coexistence equilibrium points $Q_2$ and $Q_3$ arise through a saddle-node bifurcation at a specific value of $\frr \approx 0.28$ (blue square in Fig~\ref{fr_ep_coordinates}a and second panel in Fig~\ref{fr_ep_coordinates}b). This bifurcation value splits the interval of positive $\frr$ into a subinterval in which the virus does not persist, governed by the stability of $Q_1$, and a subinterval that allows persistent infections. While $Q_2$ remains locally stable for increasing values of $\frr$, first as a stable node and then as a spiral sink in two directions of the eigenspace generated by the linearisation and a stable node in the third one (hereafter, and otherwise stated, simply spiral sink and stable node; light orange region in Fig~\ref{fr_ep_coordinates}), $Q_3$ is born as a saddle and becomes non biologically meaningful with $y$ and $V$ coordinates becoming negative (grey curve in Fig~\ref{fr_ep_coordinates}a). Right after the saddle-node bifurcation, there exists a small transition interval of $Q_1$-$Q_2$ bistability (green region in Fig~\ref{fr_ep_coordinates}) in which both extinction and persistence can be achieved depending on the initial conditions.

The persistence interval of $\frr$ is, in turn, split into four subintervals in the interval studied for increasing values of $\frr$: the aforementioned region of $Q_1$-$Q_2$ bistability (where both $Q_1$ and $Q_2$ are stable nodes), the interval governed by $Q_2$ as a local spiral sink and stable node in which populations oscillate towards it in damped oscillations, the interval governed by attractor limit cycles in which populations keep oscillating for infinite time (blue region in Fig~\ref{fr_ep_coordinates}a and fourth panel in Fig~\ref{fr_ep_coordinates}b) and, again, an interval governed by $Q_2$ as a local spiral sink and stable node. In all of them, persistent infections are established. Still, the equilibrium state reached is different: in those scenarios governed by $Q_2$, persistence is achieved by convergence of oscillatory populations of cells and virus towards a stable state; on the other hand, in the interval governed by the attractor limit cycle, waves of infection keep populations fluctuating. The transition between the two scenarios is a Hopf bifurcation (grey square in Fig~\ref{fr_ep_coordinates}a) in which $Q_2$ loses stability and transfers it to a limit cycle, or, conversely, the limit cycle loses stability and $Q_2$ regains it. Fig~\ref{fr_ep_coordinates}c displays, schematically, the stability transition of the meaningful equilibria as $\frr$ increases.

\subsubsection{Attracting limit cycles explain infection waves}
\begin{figure}[ht!]
    \centering
    \includegraphics[width=\linewidth]{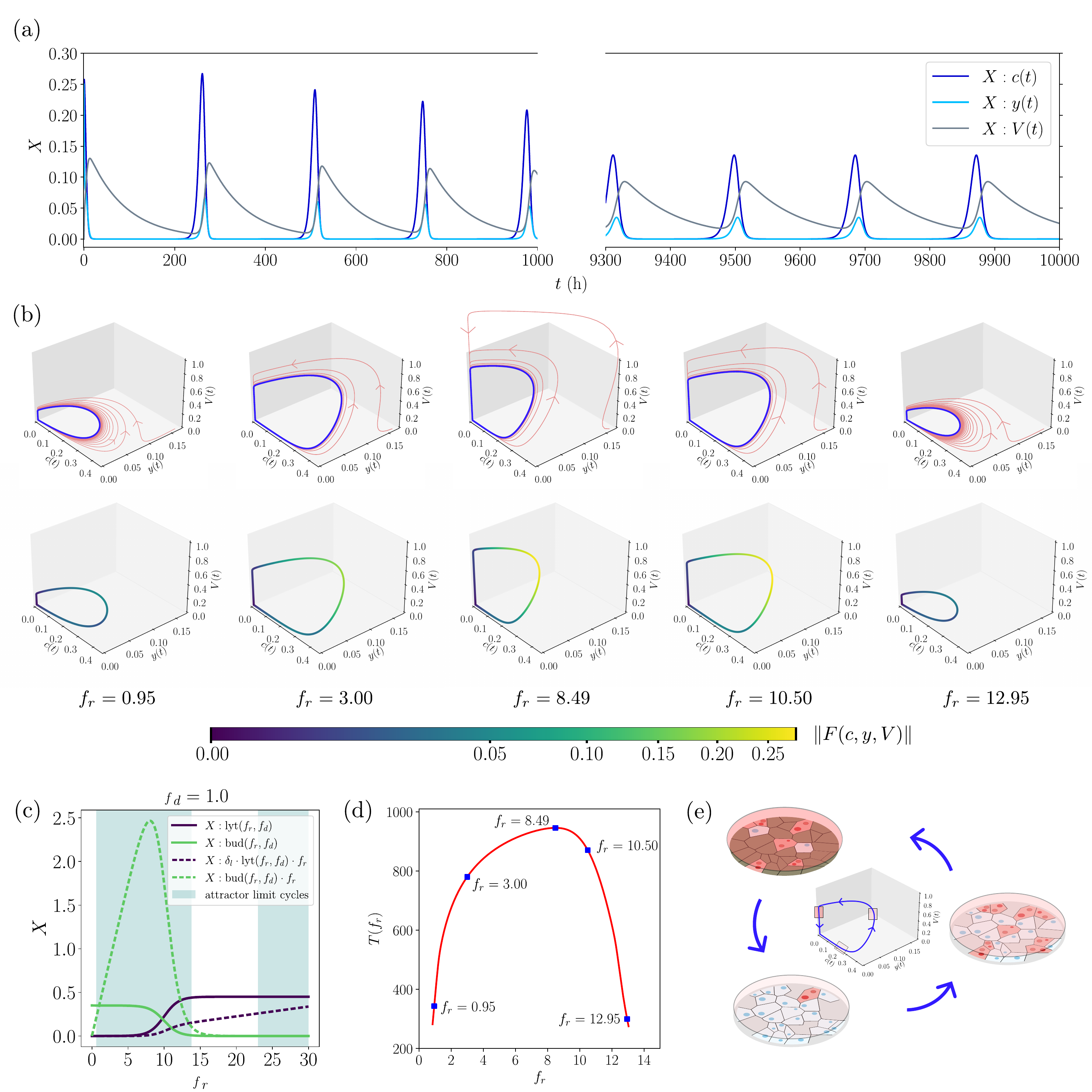}
    \caption{\textbf{Persistence achieved through oscillating populations.} (a) Time series of cell and viral populations approaching a limit cycle for $\frr = 0.7$. (b) Phase portraits of the attractor limit cycles and trajectories approaching them for five different values of $\frr$ (first row); and speed of the limit cycles at each point given by the norm of the vector field (second row). Five different limit cycles for different values of $\frr$ are displayed. (c) Contribution of the processes of lysis and budding to the population of virus in the supernatant as a function of the replicative fitness $\frr$. (d) Period of the limit cycles as a function of $\frr$. (e) Stages of the infected cell culture at different points of a limit cycle.}
    \label{LC_Figure}
\end{figure}

For a fixed $\frr$ in the interval in which the dynamics are attracted by limit cycles, the populations of cells and virus oscillate, asymptotically approaching an isolated periodic orbit. The oscillatory behaviour establishes a trade-off between populations: $V$ decreases, primarily, by infecting cells, resulting in an increasing $y$ population and decreasing density of uninfected cells $c$. As virions are released to the supernatant and infected cells die (either killed by lysis or degradation) while uninfected cells divide, $V$ and $c$ populations grow and $y$ decreases, completing an infection wave. This periodic process is illustrated in the time series in Fig~\ref{LC_Figure}a: the amplitude of the oscillations approaches that of the attractor limit cycle, which represents the oscillatory stable state of the infection over very long times.

The amplitude of the limit cycle in the variable $V$ is the maximum viral titer achieved during the course of an infection. The numerical results suggest the existence of a value of $\frr \approx 8.49$ at which the maximum amplitude is reached, as displayed in the first row of Fig~\ref{LC_Figure}b. Thus, increasing the replication fitness $\frr$ of the virus does not necessarily increase the maximum attainable viral titer. The maximal amplitude is obtained at the maximum of the function $\bud(\frr,\fdd) \cdot \frr$, which is achieved when the budding strategy still dominates over the lytic exit and the predominance shift begins (Fig~\ref{LC_Figure}c). In this sense, the added effect of $\frr$ fosters the budding exit above the lytic release: comparing the dashed curves in Fig~\ref{LC_Figure}c (green and dark violet), the effect of $\frr$ is stronger on the budding than on the lytic function. The budding strategy is optimal in terms of offspring production at a specific value of $\frr$: an increase in $\frr$ does not necessarily lead to an increase in the virions released when $\fdd$ is constant. This threshold value of $\frr$ corresponds to the optimal equilibrium between replication and dispersion, maximising the viral titer without triggering cell lysis. On the contrary, the monotonicity of $\lyt(\frr,\fdd) \cdot \frr$ (dashed, dark violet curve in Fig~\ref{LC_Figure}c) establishes the optimal strength of the lytic exit in terms of virion release at the highest attainable value of $\frr$. 

The study of the speed at which populations oscillate near the limit cycle reveals two distinct regimes: a fast one in which all populations are far from zero, and a slow one in which either the cell population or the virus approaches zero (the speed can decrease by up to six orders of magnitude depending on the value of $\frr$). The second row in Fig~\ref{LC_Figure}b displays the modulus of the vector field in the limit cycle for different values of $\frr$. The evolution of the infection lags when the cell population becomes very small, so degradation causes the viral population to decay. Cell division gradually restores the population, and the infection process reactivates. Thus, the infection evolves swiftly when there is enough concentration of both virus and cells and slows down and starts a process of slow recovery when any of the variables approaches extinction. Replication fitness $\frr$ within this range allows viral recovery despite viral populations approaching demise.

Moreover, the numerical results show that the maximum speed is attained at the optimal $\frr$ for the budding exit (central plot in the second row of Fig~\ref{LC_Figure}b), suggesting that, at this value, virus and cell populations change very fast in the way of the infection towards the maximum viral titer. The longest period of the limit cycle is also attained at this value (Fig~\ref{LC_Figure}d). Then, viral populations possessing this optimal $\frr$ trigger longer infection waves, which may cause large viral titers for longer periods. Finally, Fig~\ref{LC_Figure}e depicts the qualitative evolution of the infection in the experimental cell cultures. 

\subsection{Hierarchical effects of fitness parameters on infection dynamics}
\begin{figure}[ht!]
    \centering
    \includegraphics[width=\linewidth]{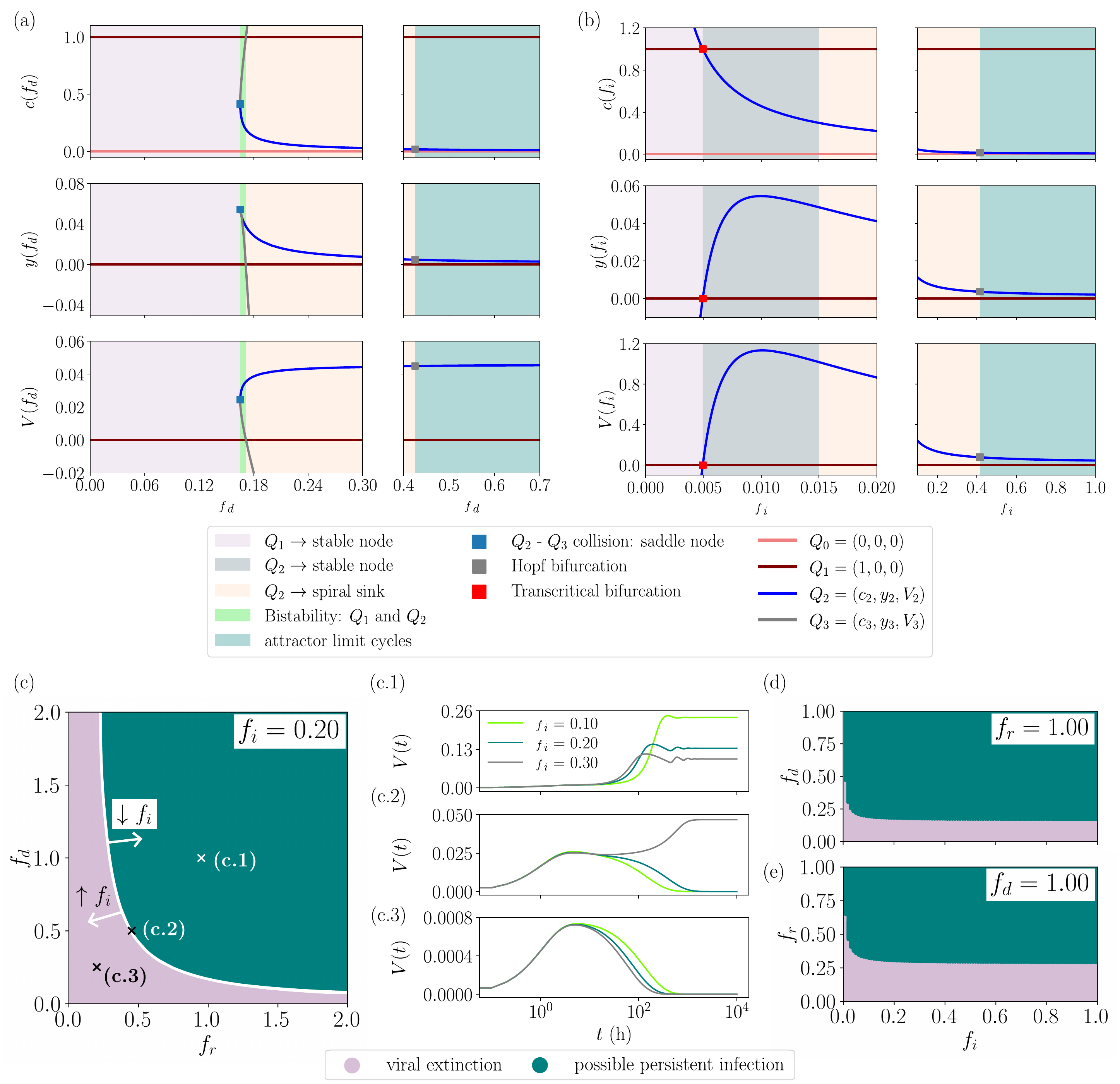}
    \caption{\textbf{Equilibrium points, local stability, and existence in $\mathcal{F}$ at increasing $\fdd$ and $\fii$.} (a) Coordinates $c$, $y$ and $V$ (top, middle and bottom panels, respectively) and local stability regions of the biologically meaningful equilibrium points as a function of $\fdd$ with constant values $\frr = \fii = 1$. (b) Same as in (a), now as a function of $\fii$ with constant values $\frr = \fdd = 1$; (c) Outcomes of the infection in terms of the local stability of the extinction equilibrium point $Q_1 = (1, 0,0)$ in the plane $(\frr, \fdd$). (d) Same as in (c), now in the plane $\fii - \fdd$. (e) Same as in (d) for the plane ($\fii - \frr$); (c.1-c.3) Time series of the viral titer computed for different values of $\fii$ at different points of the plane ($\frr, \fdd$): at the region of possible persistence, at the boundary between regions and at the region of extinction, respectively.}
    \label{fdfi_eq_points}
\end{figure}

Variations in $\fdd$ reveal an analogous evolution of the equilibrium points to that observed in variations of $\frr$ for single-passage dynamics. Low dispersal fitness values drive the infection to extinction, and there exists a saddle-node bifurcation value above which viral persistence is allowed (Fig~\ref{fdfi_eq_points}a). The interval of persistence is split into the interval governed by the stable coexistence equilibrium $Q_2$ and the interval governed by attractor limit cycles, both of them separated by a Hopf bifurcation. 

The upper bound of the extinction interval in terms of $\fii$ shrinks by two orders of magnitude (Fig~\ref{fdfi_eq_points}b). The infection demise is only achieved for very small values of $\fii$ that correspond to a virus almost unable to infect. In this case, viral degradation would drive the viral population to extinction. Therefore, a viral population with $\frr$ and $\fdd$ values within the interval of persistence would not become extinct due to changes in $\fii$ unless the virions lose their ability to infect. 

The local stability of $Q_1$ provides insight into the fitness regions in the fitness space where viral clearance drives the infection to its demise. The analogy in the evolution of the equilibrium points as a function of both $\frr$ and $\fdd$ is observed in the quasi-symmetrical pattern of Fig~\ref{fdfi_eq_points}c, in which the regions of viral extinction and viral persistence in the parameter space $(\frr, \fdd)$ are plotted. Changes in any of them may shift the outcome of the infection from persistence to extinction, or vice versa. This result is robust to changes in $\fii$, as shown in the time series of Figs~\ref{fdfi_eq_points}c.1-\ref{fdfi_eq_points}c.3. By increasing or decreasing $\fii$, the outcome of the infection can only change at the border curve between the extinction and the persistence regions. Far from it, small perturbations in $\fii$ do not alter the course of the infection, as Figs~\ref{fdfi_eq_points}c.1 and~\ref{fdfi_eq_points}c.2 show.

The symmetry in the distribution of the regions is not observed when $\fdd$ and $\frr$ are tested against $\fii$ (Figs~\ref{fdfi_eq_points}d and~\ref{fdfi_eq_points}e). Mutations causing changes in $\fii$ may not disrupt the outcome of the infection, while changes in $\frr$ or $\fdd$ may do so. Thus, these findings underscore the dominant role of $\frr$ and $\fdd$ relative to $\fii$.

\subsection{Fitness loss can drive abrupt and smooth transitions to viral extinction}

In this section, we present the results obtained by simulating multiple finite-time passages following the methodology described above. The simulated data considered here consist of a single observable at each passage, namely, the extracellular infectious viral titer $V$, whereas the model contains three potentially varying fitness components, $\frr$, $\fii$, and $\fdd$, that can not be measured in real time. Estimating these three quantities independently at every passage would therefore constitute an underdetermined inference problem, in which multiple fitness trajectories could provide similarly good descriptions of the same experimental data. In particular, reductions in replicative and dispersal fitness can produce comparable effects on extracellular viral titer and cannot be distinguished using these measurements alone. Consequently, we do not fit the fitness parameters to the experimental trajectories or interpret the values used in Table~\ref{tab:fitness} as empirical estimates. Instead, we consider six illustrative realisations showing that passage-to-passage fitness variations can qualitatively reproduce the two patterns observed experimentally: sustained fluctuations followed by an abrupt loss of detectable infectivity, or a progressive decline preceding extinction. Thus, these simulations constitute a proof of principle showing that approaching and crossing the persistence--extinction boundary can generate the observed qualitative behaviours, rather than a reconstruction of the actual fitness evolution of the experimental viral populations.

\begin{table}[!ht] 
\centering
\caption{\textbf{Illustrative fitness trajectories across passages driving abrupt and smooth transitions to viral extinction.}}
\resizebox{\linewidth}{!}{%
\begin{tabular}{|c|c|*{15}{c}|}
\hhline{~~|---------------|}
% Row 1
\multicolumn{2}{c|}{} &
\multicolumn{15}{c|}{\textbf{Passage number}} \\ 
\hhline{~~|---------------|}
% Row 2
\multicolumn{2}{c|}{} &
0 & 1 & 2 & 3 & 4 & 5 & 6 & 7 & 8 & 9 & 10 & 11 & 12 & 13 & 14 \\ \hhline{~~|---------------|} \noalign{\vskip 5pt}

% Replicate 1
\hline \multirow{3}{*}{Run 1} & $\frr$ &
1.0 & 1.0 &  1.05 &  1.1 &  1.15 &  1.20 &  1.25 &  1.3 &  1.35 &  1.4 &  1.45 &  1.5 &  1.55 &  1.6 &  1.65 \\ 
& $\fdd$ &
4.5 &  5.7 &  8.9 &  4.1 &  10.9 &  1.5 &  6.7 &  5.2 &  7.8 &  0.02 &  0.01 &  0.0 &  0.0 &  0.0 &  0.0 \\ 
& $\fii$ &
1.5 &  3.6 &  8.9 &  4.3 &  2.6 &  5.4 &  1.2 &  7.8 &  3.4 &  0.10 &  0.01 &  0.0 &  0.0 &  0.1 &  0.2 \\ \hline \noalign{\vskip 5pt}

% Replicate 2
\hline \multirow{3}{*}{Run 2} & $\frr$ &
2.14 &  11.82 &  6.37 &  13.45 &  0.91 &  8.76 &  4.52 &  9.11 &  1.67 &  12.03 &  1.00 &  7.29 &  10.64 &  3.40 &  13.02 \\ 
& $\fdd$ &
3.52 &  11.87 &  0.44 &  9.13 &  6.71 &  13.02 &  0.10 &  0.06 &  0.01 &  0.002 &  0.001 &  0.028 &  0.009 &  0.073 &  0.144 \\ 
& $\fii$ &
8.42 &  1.77 &  13.21 &  5.66 &  10.03 &  0.91 &  7.48 &  12.84 &  3.19 &  0.001 &  0.0041 &  0.0093 &  0.0007 &  0.0065 &  0.0028 \\ \hline \noalign{\vskip 5pt}

% Replicate 3
\hline \multirow{3}{*}{Run 3} & $\frr$ &
2.14 &  11.82 &  6.37 &  13.45 &  0.91 &  8.76 &  4.52 &  9.11 &  1.67 &  12.03 &  1.00 &  7.29 &  10.64 &  3.40 &  13.02 \\ 
& $\fdd$ &
2.5 &  2.1 &  1.8 &  1.4 &  0.9 &  0.6 &  0.2 &  3.4 &  0.099 &  0.065 &  0.065 &  0.01 &  0.028 &  0.009 &  0.073 \\ 
& $\fii$ &
2.5 & 2.7 & 2.9 & 3.1 & 3.3 & 3.5 & 3.7 & 3.9 & 4.1 & 4.3 & 4.5 & 4.7 & 4.9 & 5.1 & 5.3 \\ \hline \noalign{\vskip 5pt}

% Replicate 4
\hline \multirow{3}{*}{Run 4} & $\frr$ &
1.0 &  1.0 & 1.0 & 1.0 & 1.0 & 1.1 & 1.1 & 1.0 & 1.0 & 1.0 & 1.0 & 1.0 & 1.0 & 1.0 & 1.0 \\ 
& $\fdd$ &
2.5 &  10.5 &  1.8 &  4.5 &  0.1678 &  0.16779 &  0.16778 &  0.167779 &  0.167778 &  0.167779 &  0.160 &  0.065 &  0.01 &  0.028 &  0.009 \\ 
& $\fii$ &
1.0 & 1.0 & 1.0 & 1.0 & 1.0 & 1.0 & 1.0 & 1.0 & 1.0 & 1.0 & 1.0 & 1.0 & 1.0 & 1.0 & 1.0 \\ \hline \noalign{\vskip 5pt}

% Replicate 5
\hline \multirow{3}{*}{Run 5} & $\frr$ &
2.14 &  11.82 &  6.37 &  13.45 &  0.91 &  8.76 &  4.52 &  4.50 &  4.49 &  4.55 &  1.00 &  7.29 &  10.64 &  3.40 &  13.02 \\ 
& $\fdd$ &
3.52 &  11.87 &  0.44 &  9.13 &  6.71 &  13.02 &  0.035 &  0.034999 &  0.034998 &  0.034997 &  0.0329 &  0.028 &  0.009 &  0.073 &  0.144 \\ 
& $\fii$ &
8.42 &  1.77 &  13.21 &  5.66 &  10.03 &  0.91 &  7.48 &  7.49 &  7.48 &  7.46 &  0.0041 &  0.0093 &  0.0007 &  0.0065 &  0.0028 \\ \hline \noalign{\vskip 5pt}

% Replicate 6
\hline \multirow{3}{*}{Run 6} & $\frr$ &
12.3 & 12.3 & 12.3 & 12.3 & 12.3 & 12.3 & 12.3 & 12.3 & 12.3 & 12.3 & 12.3 & 12.3 & 12.3 & 12.3 & 12.3 \\ 
& $\fdd$ &
10.54 &  1.53 &  10.56 &  3.57 &  2.78 &  9.8 &  1.03 &  0.152 &  0.16700 &  0.17900 &  0.17900 &  0.11997 &  0.028 &  0.009 &  0.073 \\ 
& $\fii$ &
10.67 &  1.77 &  2.78 &  13.21 &  5.66 &  10.03 &  0.91 &  7.48 &  7.48 &  7.48 &  7.48 &  0.0041 &  0.0093 &  0.0007 &  0.0065 \\ \hline

\end{tabular}}
\begin{flushleft} 
Values of \(f_r\), \(f_d\), and \(f_i\) used in the six simulated realizations shown in Fig. 7. These values were not estimated from the experimental data and should not be interpreted as empirical fitness measurements. We selected them to illustrate how different passage-to-passage trajectories through fitness space can generate  (runs 1-3) or progressive (runs 4-6) transitions from viral persistence to extinction.
\end{flushleft}
\label{tab:fitness}
\end{table}

\begin{figure}[ht!]
    \centering
    \includegraphics[width = \linewidth]{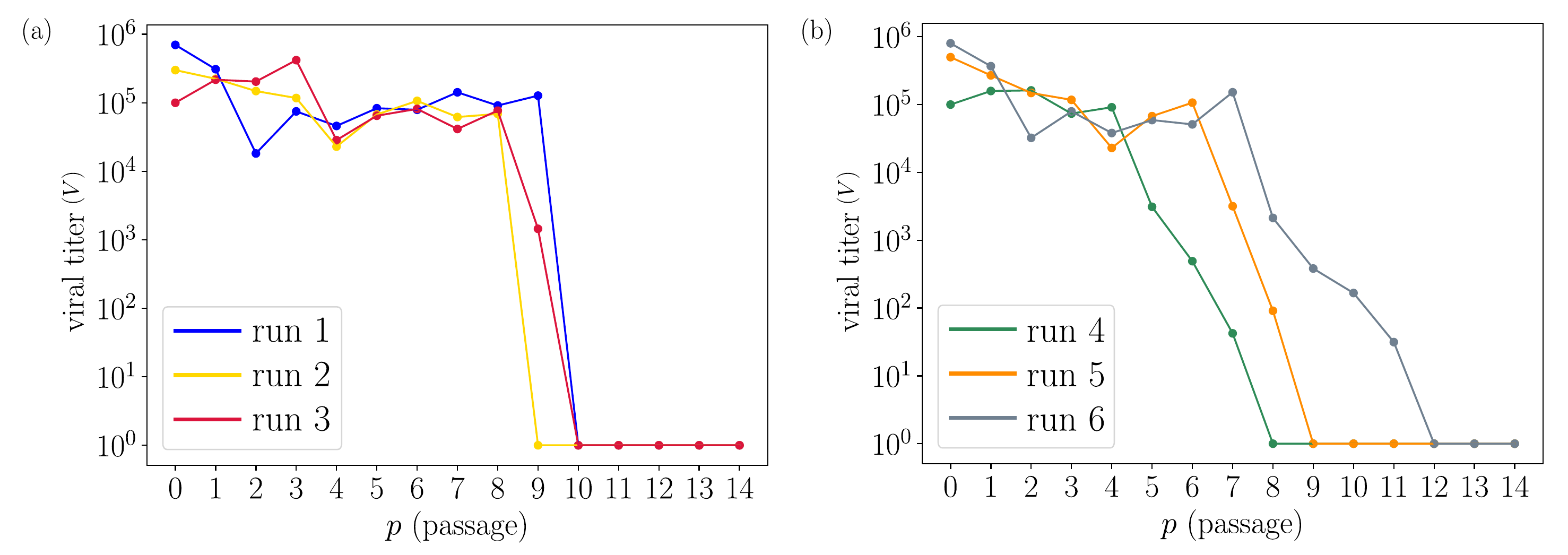}
    \caption{\textbf{Simulated viral dynamics across passages.} Viral titer at the end of each passage for six model realisations exhibiting (a) abrupt and (b) progressive transitions to viral extinction. The passage-dependent values of $f_r$, $f_i$, and $f_d$ are given in Table~2. These values are illustrative and were not estimated from the
    experimental data. Viral populations are normalised by the carrying capacity $K$.}
    \label{fig:passages}
\end{figure}

Figure~\ref{fig:passages} displays the viral population in the supernatant $V$ (viral titer) at the end of each passage for the six simulated realisations. Each realisation corresponds to a different initial condition for the uninfected-cell and viral populations in the first passage and to a different sequence of fitness values over the subsequent passages. The viral titers simulated in the three realisations shown in Fig.~\ref{fig:passages}a fluctuate until passages 8 or 9, depending on the realisation, and then undergo a sharp decline leading to the termination of the infection. Until this point, the viral population has fitness values within the persistence region, where either the coexistence equilibrium $Q_2$ or an attracting limit cycle governs the dynamics. Because each passage is interrupted after 72 hours, the measured viral titer corresponds to a particular stage of the transient oscillation towards $Q_2$ or of the oscillation around the limit cycle, depending on the fitness values. Consequently, the titer reached at the end of a passage depends not only on the fitness parameters, but also on the initial condition. When either $\frr$ or $\fdd$ crosses the saddle-node bifurcation threshold and enters the extinction region, the viral population collapses, shifting the system from persistence to viral clearance.

The simulated realisations 4, 5, and 6, shown in Fig.~\ref{fig:passages}b, differ from those in Fig.~\ref{fig:passages}a in that the viral titer declines progressively over several passages. In these examples, either $\frr$ or $\fdd$ approaches the saddle-node bifurcation value, causing a slowing down of the dynamics \cite{Kuznetsov1998ElementsEdition}. As the system comes near this threshold, increasingly long times are required for the viral population to move towards the persistent state. Because each passage starts with $V=0$ and is terminated after 72 hours, this slowing down results in progressively lower viral titers at the end of successive passages. The viral population remains close to the boundary between persistence and extinction until a subsequent fitness change moves the system into the extinction region.

These illustrative simulations therefore show that passage-to-passage fitness changes that drive the viral population towards sufficiently low values of $\frr$ or $\fdd$ can result in the loss of persistence. Moreover, the trajectory followed relative to the saddle-node bifurcation determines the qualitative form of the decline: a direct crossing of the threshold results in an abrupt collapse, whereas a gradual approach to the threshold produces a progressive decrease in viral titer before extinction.

\section{Discussion}
In this work, we present a mathematical framework that highlights the role of viral fitness in the course of a persistent infection in cell culture. By placing viral fitness at the core of the mathematical model described, we aim to assess the importance of this viral trait in maintaining persistently infected cells. Some models have previously addressed viral persistence by considering the host immune response and emphasising competition dynamics between viral particles and immune cells \cite{Browne2018DynamicsNetwork, Ganusov2013MathematicalResponses}. Although we do not include the action of the immune response explicitly in our model, some parameters also capture this effect on persistence maintenance. For example, a decrease in free viral particles mimics neutralising antibody activity, and a decrease in infected cells mimics CTL activity. However, we give full attention to how infectious viral particles carry out a full replication cycle that depends directly on the phenotypic abilities acquired through mutations. These capacities include replication, dispersal and infection of cells. By identifying these processes as the main actions that all infectious virions must be able to perform, we have studied, using a mathematical model, how changes in these capacities steer the virus along distinct evolutionary pathways affecting persistence. We have associated each aforementioned viral trait with a different fitness: replicative fitness $\frr$, dispersal fitness $\fdd$ and infective fitness $\fii$, respectively. In this way, we extend the meaning of viral fitness beyond the experimentally measured replicative fitness. \\

Based on the experimental data (Fig~\ref{fig:exp:data}) of how two distinct HCV populations with different replicative fitness \cite{Moreno2017InternalEnvironment} evolve across passages, our model incorporates an explicit, continuous dependence on $\frr$ of the lytic behaviour of the viral particles. Contrary to other models of chronic infections that include the lytic and budding strategies \cite{Gulbudak2019HeterogeneousSystems, Gulbudak2021ACells}, viruses in our model are not assumed to have a fixed pattern to exit the cell, but an evolving one. Therefore, a viral population may release new virions into the supernatant either by cell bursting or by budding through the cell membrane, depending on the replicative fitness acquired through mutations. This explicit dependence on replicative fitness was observed in both HCV p0 (low replicative fitness) and HCV p200 (high replicative fitness) populations described in \cite{Moreno2017InternalEnvironment}. Not all cell exit strategies are necessarily inherent to a viral population, but each one may dominate to an extent that depends on the viral replicative ability. A low replicative fitness HCV population that killed cells but produced infectious virions was unable to establish a persistent infection. Some other models have made approximations to the definition of new fitness parameters, but from a within-cell perspective~\cite{Weitz2019ViralLatency}. Other models do not incorporate the different cell exit strategies explicitly as a continuum ~\cite{Khan2024DiscreteBifurcations}. \\

Hence, by introducing three different types of fitness, our model increases the dimensions of the space: from the original 3-dimensional space of variables $c$ (uninfected cells), $y$ (infected cells) and $V$ (viral particles in the cell culture supernatant) to a 6-dimensional space, adding the replicative fitness $\frr$, the dispersal fitness $\fdd$ and the infective fitness $\fii$ as new degrees of freedom of the system. However, we do not present a mechanistic description of the 3-dimensional space of fitness. The difficulty of measuring viral fitness during an ongoing infection, and consequently, its evolution in time, strengthens the necessity of a theoretical framework that tries to disclose the role of viral fitness in establishing persistent infections. Rather, we have tackled the problem in terms of fitness gains and losses as a first approach to the mutation strategies that lead the virus to modify its fitness. To do so, we have sampled different values of the fitness parameters. By distinguishing among the three basic abilities of a virus and identifying each of them with a different fitness allows deepening the understanding of how evolutionary mechanisms such as mutation or selection can drive an infection towards different stages in the absence of the host immune response. By doing so, multiple new viral strategies are unveiled since fitness is not restricted to the replicative capacity.  \\

We first focus on the fate of the infection, simulating a single, long-term passage experiment. The bifurcation analysis with respect to the replicative fitness $\frr$ (for constant values of $\fdd$ and $\fii$) reveals two main outcomes: total extinction of the virus present in the cell culture medium, and the establishment of a persistent infection due to the virus survival in the culture medium. These two scenarios are separated by a saddle-node bifurcation at a critical value $\frr^*$ (blue square in Fig~\ref{fr_ep_coordinates}a). The emergence of a coexistence equilibrium point as a consequence of the bifurcation allows the viral population to maintain a non-zero viral titer over time. Within the persistence regime (for $\frr > \frr^*$), the viral population either converges to a coexistence equilibrium through damped oscillations, or keeps oscillating in a limit cycle (Fig~\ref{LC_Figure}a and Fig~\ref{LC_Figure}b). In terms of $\frr > \frr^*$, persistence can therefore be achieved through these two distinct dynamical regimes, separated by a Hopf bifurcation. An analogous scenario emerges for the dispersal fitness $\fdd$ (Fig~\ref{fdfi_eq_points}a). In contrast, only negligible values of the infective fitness $\fii$ drive the virus towards extinction (Fig~\ref{fdfi_eq_points}b). Thus, our model reveals a hierarchy within the three-dimensional fitness space: while variations in $\frr$ and $\fdd$ can shift the outcome of the infection between extinction and persistence, variations in $\fii$ alone are unable to modify the outcome (Fig~\ref{fdfi_eq_points}c). 

Within the region of limit cycles, we have identified a value of $\frr$ at which the maximum viral titer is attained, corresponding to the maximum amplitude of the limit cycle. At this value, the budding strategy is optimal in terms of offspring production (Fig~\ref{LC_Figure}c). Beyond this $\frr$ value, the virus becomes gradually lytic and combines both cell exit strategies. Our model shows that the coexistence of both exit strategies at similar levels is less effective for the virus to obtain high viral titers in the course of the infection. When the virus becomes mainly lytic, higher $\frr$ values lead to higher viral titers attained in the oscillatory viral populations, as opposed to what is observed when the virus is mainly non-lytic (the budding strategy dominates). Indeed, only a fraction $\delta_{\ell}$ of killed infected cells releases virions when the viral population is mainly lytic, whereas all infected cells contribute to the release of virions to the supernatant when the viral population is mainly non-lytic (for equal values of $\fii$). Consequently, achieving higher viral titers in the lytic phase necessitates significantly larger $\frr$ values compared to the non-lytic phase. We have identified, in terms of viral titer, that the budding strategy becomes optimal for constant division rates of susceptible cells. According to \cite{Weitz2019ViralLatency}, this result suggests that cells do not grow rapidly enough so that viruses keep their hosts alive to ensure their persistence as an optimal strategy. The scarcity of susceptible cells that leads the viral population to adopt the budding strategy to achieve higher viral titers is well illustrated in the time series in Fig~\ref{LC_Figure}a, where drops in the $c$ population, maintained over long periods, are observed. \\

The oscillations towards the stable state of equilibrium (either a coexistence equilibrium point or a limit cycle) in the persistence region are the basis of the oscillations in viral titer that are measured at the end of a passage. Finite-time predefined passages lead to interrupting the course of the infection on its way towards a stable state. Oscillating viral titers within this regime depend on the initial proportion of infected versus uninfected cells in the cell culture at the beginning of the passage. Therefore, different initial proportions of cells lead to different viral titers at the end of the passage, causing oscillations in the extracellular viral titer across passages (Fig~\ref{fig:passages}). In line with the oscillatory extracellular viral titer observed experimentally for the HCV before the loss of persistence (Fig~\ref{fig:exp:data}), similar patterns of oscillating viral titers have been observed in persistently infected cells with the infectious pancreatic necrosis virus (IPNV)~\cite{Chaumont2026CharacterizationVirus} and the foot-and-mouth disease virus (FMDV)~\cite{Herrera2008PersistenceEvolution}. Our model offers a mechanistic description of these oscillatory behaviours in terms of viral fitness. The existence of the equilibrium point $Q_2$ and a Hopf bifurcation within the fitness region of persistence is critical to observing long-term oscillations during a passage that induce the fluctuations of the viral titer in the course of all passages. \\

Changes in replicative and dispersal fitness could trigger the demise of the infection. The experimental results show a viral titer drop for both HCV populations (p0 in Fig~\ref{fig:exp:data}a and p200 in Fig~\ref{fig:exp:data}). Our model shows that comparable drops can arise when either replicative or dispersal fitness decreases sufficiently to move the system from the persistence region towards viral clearance. Given the aforementioned hierarchy in fitness space, only $\frr$ and $\fdd$ split the interval of attainable fitness values into two regions: extinction and persistence. The crossing of the saddle-node bifurcation boundary due to a decrease in $\frr$ or $\fdd$ shifts the outcome of a passage from persistence to extinction (drop of viral titer in replicates of Fig~\ref{fig:passages}). As opposed to other mathematical models that identify cell traits as key parameters to ensure viral persistence \cite{FRANK2000Within-hostParticles}, our results emphasise the importance of viral fitness in establishing persistent infections in cell cultures. The slowing down of the dynamics close to a saddle-node bifurcation~\cite{Strogatz2024NonlinearEdition} causes the gradual decline of the viral titer observed in Fig~\ref{fig:passages}b. Indeed, a viral population that loses either $\frr$ or $\fdd$ and approaches the bifurcation value will slow down the dynamics towards persistence, inducing longer times to reach non-zero viral titers. Therefore, the interruption of the infection to measure the viral titer and passage cells will occur at a lower viral load $V$ the nearer the system gets to the bifurcation point. This phenomenon causes a smooth decline of the viral titer across passages until extinction. The shift from the persistence region to the extinction region without taking fitness values close to the bifurcation point will cause a sharp decrease in the viral titer (Fig~\ref{fig:passages}a). As a result of this, the distance of the viral fitness to the saddle-node bifurcation value at which the viral population shifts from persistence to extinction determines the nature of the viral titer decay.

Besides the extinction mechanisms based on shifts of viral fitness, our model also contemplates drops of the viral titer due to the interruption of the course of the infection in the context of finite-time experimental passages. Indeed, limit cycles in the persistence region of $\frr$ or $\fdd$ that approach $V \approx 0$ may induce undetectable viral titers if the infection is halted at that time point. Thus, the experimental design and, more importantly, the viral and cell traits (they shape the limit cycles and determine whether the oscillations approach zero during an infection) also play a crucial role in determining the outcome of a multi-passage experiment.

Although the differences in the viral titer decay observed experimentally (Fig~\ref{fig:exp:data}a and Fig~\ref{fig:exp:data}b) are associated with the difference in the initial replicative fitness of the HCV populations, we stress that our model reproduces both dynamics without the need to take into account the initial value of $\frr$ of the population. That is, since the type of decay depends only on the way mutations drive the viral fitness of the population ($\frr$ or $\fdd$) towards the extinction region, an initial viral population with a specific $\frr$ does not determine the path towards extinction according to our model. Incorporating the coevolution of cells and the virus into the model, the next step in our research line, could shed light on the underlying mechanisms and the role of the initial $\frr$ of the viral population on the loss of persistence. Thus, the mathematical framework at the core of this work reproduces multiple strategies (fitness variation pathways) towards the same final state of the infection. Further research is needed to identify those viral or cell traits that drive viral fitness through a specific pathway in the space of fitness towards loss of persistence. \\

In conclusion, we present a new mathematical framework to help understand viral persistence, a salient problem in viral population dynamics and medical virology, based on an extended interpretation of viral fitness. \\

\begin{comment}
\textbf{També per a future work:}
\begin{itemize}
    \item Prèviament s'ha vist que les fluctuacions poden ser degudes a la competició amb DIs que es produeixen durant la infecció. No obstant, això s'ha vist en experiments en què es passen virus, no cèl·lules \cite{Palma1974CyclicParticles}. Es podrien incorporar DIs al model i veure quin efecte tenen.
    \item A \cite{FRANK2000Within-hostParticles} es presenta un model sense DI previ al model amb DI en què s'aconsegueix persistència amb fluctuacions. És un model amb espai. La principal diferència amb el model amb DIs és deguda a l'espai.
    \item In \cite{Maslov2015Well-temperateCollapses}, ja es contempla la possibilitat de transició entre estratègies de lisogènia i lisi en una mateixa població viral. S'estudia des del punt de vista dels bacteriòfags.
\end{itemize}
\end{comment}

\section*{Acknowledgments}
We want to thank Marc Jorba for stimulating discussions and technical guidance on the Taylor high-accuracy ODE numerical integrator. We also thank Carlos García-Crespo for his valuable support during the initial setup of the experiments. This work was supported by the Spanish Ministry of Science and Innovation, grants PID2020-113888RB-I00/AEI/10.13039/501100011033 (to E.D.) and 202220I116 (to C.P.), by the Spanish Ministry of Science, Innovation and Universities (MICIU), grant PID2023-146622OB-I00 (to C.P. and E.D.) financed by MICIU/AEI/10.13039/501100011033, by FEDER, UE, through the “Severo Ochoa” Programs for Centers of Excellence in R\&D [CEX2021-001154-S], and by the European Commission-Next Generation EU (regulation EU 2020/2024) through the CSIC’s Global Health Platform (PTI Salud Global). We also acknowledge the project TEC-2024/BIO-66 (SALAINDEC-CM from Comunidad de Madrid/FEDER). We also acknowledge institutional grants from the Fundación Ramón Areces and Banco Santander to the CBM. The CBM team belongs to the Global Virus Network (GVN) and the Biomedicine Unit of Universidad de Castilla-La Mancha (UCLM), associated with CSIC. J. Tomás Lázaro has also been funded by Grant No. PID2024-155942NB-I00 MICIU/AEI/ and "ERDF a way of making Europe". He was also supported by the Spanish State Research Agency, through the Severo Ochoa and María de Maeztu Program for Centers and Units of Excellence in R\&D (CEX2020-001084-M) (JS and JTL). We thank CERCA Programme/ Generalitat de Catalunya for institutional support.

\begin{comment}
\section*{Author contributions}
\begin{itemize}
    \item Conceptualization: Oriol Llopis-Almela, J. Tomás Lázaro, Esteban Domingo, Celia Perales and Josep Sardanyés

    \item Data curation: Antoni Duran, Celia Perales, Esteban Domingo

    \item Formal analysis: Oriol Llopis-Almela, J. Tomás Lázaro, Antoni Duran, Josep Sardanyés

    \item Funding acquisition: J. Tomás Lázaro, Esteban Domingo, Celia Perales and Josep Sardanyés

    \item Investigation: Oriol Llopis-Almela, J. Tomás Lázaro, Esteban Domingo, Antoni Duran, Celia Perales and Josep Sardanyés

    \item Methodology: Oriol Llopis-Almela, J. Tomás Lázaro, Esteban Domingo, Antoni Duran, Celia Perales and Josep Sardanyés

    \item Software: Oriol Llopis-Almela

    \item Supervision: J. Tomás Lázaro, Esteban Domingo, Celia Perales and Josep Sardanyés

    \item Writing – original draft: Oriol Llopis-Almela

    \item Writing – review and editing: Oriol Llopis-Almela, J. Tomás Lázaro, Esteban Domingo, Celia Perales and Josep Sardanyés
\end{itemize}
\end{comment}

%\nolinenumbers
\bibliography{references}

\newpage
%\linenumbers
\section*{Appendix}
\begin{appendices}
\section{Model parameters}\label{AppA}
The table below contains the values used in the numerical simulations for some parameters of the model and their justification, based on the experimental evidence of HCV infections in cell cultures.

\begin{longtable}[h!]{|p{0.13\linewidth}|p{0.08\linewidth}|p{0.77\linewidth}|} \hline
\textbf{Parameter} & \textbf{Value} & \textbf{Justification} \\ \hline
$\alpha_c$ & 0.023 hours$^{-1}$ & The population of uninfected cells duplicates every 24--36 hours. Assuming an exponential growth $c(t) = c_0 \exp(\alpha_c t)$ in the absence of infected cells and viruses and without limitations of space or resources, $\alpha_c$ is given by the well-known doubling time law:
\[
    \alpha_c = \frac{\ln 2}{\textrm{doubling time}}
\]
Therefore, given the range 24--36 hours for the doubling time, we have $\alpha_c \in [\ln 2 / 36, \ln 2 / 24] \approx [0.019, 0.029]$. \\ \hline

 $\alpha_y$ & 0.012 hours$^{-1}$ & The doubling time of the population of infected cells is estimated to be between 1.5 and 2 times longer than that of the uninfected cells. Thus, the population doubles every 48--72 hours under an active infection. An analogous reasoning to the $\alpha_c$ case renders $\alpha_y \in [\ln 2 / 72, \ln 2 / 48] \approx [0.0096, 0.014]$. \\ \hline

 $\kappa$ & $10^{-2}$ & There is no data available on the value of $\kappa$ in \cite{Moreno2017InternalEnvironment}, but a very low proportion of infected cells is expected to give rise to uninfected cells during cell division. We consider that only 1\% of infected cells undergo this effect, $\kappa = 0.01$. \\ \hline

 $\delta_{\ell}$ & 0.025 hours$^{-1}$ & In previous mathematical models describing the HCV dynamics \cite{Neumann1998HepatitisTherapy, Dahari2007ModelingEfficacy}, the number of cells killed by lytic processes has been estimated to be approximately within a range of $0.14 - 0.68$ cells per day, which is equivalent to a range of $0.0058 - 0.028$ cells per hour. \\ \hline

 % $\delta_d$ & $10^{-1}$ hours$^{-1}$ & The natural degradation rate of Huh-7 cells i \textcolor{red}{The degradation rate of IC covers a wide range of values since it depends on multiple factors such as the number of virions infecting a single cell or the fitness of the virus, which are the object of study of this work. Then, a large range of values must be explored.} \\ \hline

 $b_{\ell}$, $b_b$ & 0.9, 0.7 & We assume that a larger proportion of maximum virions released can be achieved via lysis ($b_\ell$) than via budding ($b_b$) due to the bursting nature of the former. \\ \hline

 % $\gamma$ & $10^{-6}$ & \textcolor{red}{Completely arbitrary value for $\gamma$. We need this order of magnitude to counteract the numerical effect of $K$.} \\ \hline

 $K$ & $10^6$ & Persistence experiments for HCV are carried out in p35 plates, which may contain, approximately, $10^6$ cells. \\ \hline

 $\varepsilon$ & $0.0123$ hours$^{-1}$ & Viral titer decay was fitted to an exponential curve $V(t) = V_0 \cdot e^{-\varepsilon t}$. This equation is a solution of the ordinary differential equation $dV/dt = -\varepsilon V$. Here, $\varepsilon$ is the thermal inactivation rate constant of the infectious virus incubated at a certain temperature. For the HCV case, the values of $\varepsilon$ were estimated in~\cite{Moreno2017InternalEnvironment} at 45 $^{\circ}$C for the HCV p0 and HCV p200 with a half-life time of 1 hour. Values of $\varepsilon = 0.0125 \pm 0.0015$ (1/hours) and $\varepsilon = 0.0122 \pm 0.0014$ (1/hours) were obtained, respectively. Given that both values are within the uncertainty intervals, we take $\varepsilon = 0.0123$ as a reference value. \\ \hline

 $c_1$ & $e^{10}$ & The sigmoid constant is given by $c_1 = e^L$, which is the horizontal position of the sigmoid. That is, it controls the position of the transition region between exit strategies as a function of $\frr$. The nature of the fitness parameters is theoretical, so this value will change depending on the HCV population of reference. \\
\hline
\caption{\textbf{Model parameters.} The values provided in this table are given by the experimental evidence in \cite{Moreno2017InternalEnvironment}.}

\end{longtable}
\end{appendices}

\end{document}